\documentclass{comnet-mod}

\usepackage{amsmath} 
\usepackage{amsthm} 
\usepackage{graphics} 
\usepackage{algorithmic} 
\usepackage{subfig}
\usepackage{xcolor}
\usepackage{graphicx}
\usepackage{url}
\usepackage{float}

\begin{document}

\title{Mixed Logit Models and Network Formation}

\shorttitle{Mixed Logit Models and Network Formation} 
\shortauthorlist{Gupta and Porter} 

\author{
\name{Harsh Gupta$^*$}
\address{Department of Economics, Stanford University,\\ Stanford, CA 94305, USA\email{$^*$Corresponding author: hgupta13@stanford.edu}}
\name{Mason A. Porter}
\address{Department of Mathematics, University of California, Los Angeles,\\
 Los Angeles, CA 90095, USA; Santa Fe Institute, Santa Fe, NM 87501, USA}
}

\maketitle

\begin{abstract}
{The study of network formation is pervasive in economics, sociology, and many other fields. In this paper, we model network formation as a `choice' that is made by nodes in a network to connect to other nodes. We study these `choices' using discrete-choice models, in which an agent chooses between two or more discrete alternatives. We employ the `repeated-choice' (RC) model to study network formation. We argue that the RC model overcomes important limitations of the multinomial logit (MNL) model, which gives one framework for studying network formation, and that it is well-suited to study network formation. We also illustrate how to use the RC model to accurately study network formation using both synthetic and real-world networks. Using edge-independent synthetic networks, we also compare the performance of the MNL model and the RC model. We find that the RC model estimates the data-generation process of our synthetic networks more accurately than the MNL model. In a patent citation network, which forms sequentially, we present a case study of a qualitatively interesting scenario --- the fact that new patents are more likely to cite older, more cited, and similar patents --- for which employing the RC model yields interesting insights.}

{network formation, discrete-choice models, logit models} \medskip \medskip

2020 Math Subject Classification: 05C82, 91D30
\end{abstract}



\section{Introduction}
\label{sec:Introduction}

The analysis of network formation is a prominent topic in a diverse set of areas \cite{albert2002statistical,goldenberg2010survey,blume2011identification,graham2015methods,chandrasekhar2016econometrics}.
For example, an important topic in sociology is how homophily affects the behavior of individuals in a network \cite{socialanalysisbook,pearson2006homophily}. There are studies in economics about how network structure affects risk-sharing \cite{fafchamps2007risk}. In political science, network formation has also been used to analyze collective decision-making \cite{siegel2009social}. Researchers have also examined how the formation of sexual networks affects the transmission of sexually-transmitted diseases \cite{handcock2004likelihood}. Analyzing the mechanisms of network formation is helpful for understanding different features of a network. In particular, it can yield insights into which factors affect edge formation and network growth. 

In the present paper, we take the conceptual view that edge formation between nodes (which represent entities in a network) result from `choices' that are made by those nodes. For example, in a directed network, if there is an edge from node $i$ to node $j$, we suppose that node $i$ has chosen to connect to node $j$ (and perhaps may have chosen not to connect to other nodes in the network). This conceptual framework allows us to employ \emph{discrete-choice} (DC) models to study network formation \cite{Train_Discrete_Book}. One can use a DC model to analyze the decisions that are made by an agent (or other entity) when choosing between two or more discrete alternatives. DC models have several beneficial properties for studying network formation. First, they are flexible and can model several types of growing networks. Moreover, given a real-world network, one can study the relative {statistical accuracy} of different network-formation models. Depending on the application at hand, one can also combine multiple network-formation models. Second, using tools like logistic regression, one can readily apply DC models to real-world networks. Third, there is a well-developed literature on DC models in economics and statistics that provides tools for inference, testing, and parameter selection \cite{HosmerLemesbow,kleinbaum2002logistic,menard2002applied,Train_Discrete_Book}. We overview DC models in Section \ref{sec:Discrete-Choice-Models}.

Several econometric and sociological studies have employed DC models for empirical estimation of parameters \cite{chandrasekhar2016econometrics, snijders2010introduction, snijders2001statistical}. Overgoor \emph{et al.} \cite{DiscreteChoice} recently reported the first use (to the best of our knowledge) of DC models to directly study network formation. These studies (including the one by Overgoor \emph{et al.} \cite{DiscreteChoice}) have focused largely on the \emph{multinomial logit} (MNL) model.

The MNL model is one of the most popular DC models \cite{Train_Discrete_Book}. Although the MNL model has been extended in various ways \cite{Train_Discrete_Book, kleinbaum2002logistic}, almost all applications of it to network formation use the original MNL model. 
However, some recent work does use other MNL models for such applications. For example, Overgoor \emph{et al.} \cite{DiscreteChoice} briefly discussed the potential use of mixed-logit (MX) models (which generalize the MNL model \cite{Train_Discrete_Book}) to study network formation. 
Additionally, Tomlinson and Benson \cite{tomlinson2020learning} studied network growth using an extension of the MNL model in which the set of available options influences a chooser's relative preferences.

The original MNL model has certain features that are not well-suited to studying network formation. It assumes that each node makes exactly one choice (so it has an out-degree of $1$) and that each node has the same choice set.\footnote{We use the terms `set of alternatives', `alternative set', and `choice set' interchangeably.} These assumptions are almost never true
for real-world networks \cite{newman2010networks}. The violation of these assumptions implies that estimates from the MNL model are biased, unless the preferences are homogeneous.\footnote{The qualitative importance of this bias is different for different networks. In Sections \ref{sec:Synthetic-Network} and \ref{sec:Citation Network}, we  examine case studies in which this bias is qualitatively important.} 
We discuss these limitations of the MNL model in detail in Section \ref{subsec:LimitationsMNL}.

To overcome these limitations of the MNL model, we use a type of MX model that is known as the \emph{repeated-choice (RC)} model \cite{TrainRevelt97mixedlogit}. 
In the RC model, each node can make multiple choices, so nodes can have out-degrees that are larger than $1$. Additionally, different nodes can have different choice sets and different out-degrees. These properties are helpful when studying network growth.
We use the RC model to examine statistical associations of various network features with network formation.

We do two case studies of the RC model. In our first case study, we apply the RC model to a set of synthetic networks. We find that the RC model is able to accurately estimate the data-generation process of these synthetic networks. For these networks, we show that the RC model is accurate when the preferences of the nodes are either homogeneous or heterogeneous, whereas the MNL model is accurate only when the preferences are homogeneous.
{In our second case study, we apply the RC model to
a large patent citation network. In this case study, we face computational constraints because of
the large size of the network, so we discuss sampling methods to help
overcome such constraints. We perform robustness tests
to examine the validity of our results under different sampling methods, with which we obtain
similar qualitative results. We also perform goodness-of-fit tests to illustrate that the RC model produces networks with statistical properties that are similar to those of the parent citation network. 
We also apply the MNL model (which ignores preference heterogeneity)
to the patent citation network.
{In comparison to the MNL model, we find
that the RC model yields a better understanding of how previous citations, age, and the technology classification
of a patent affects the probability that a patent will be cited by future patents.


\subsection{Contributions and Model Applicability}

The main contribution of the present paper is to describe, employ, and illustrate the utility of the RC model in the context of network formation. To the best of our knowledge, the RC model has not been used previously to study network formation.

To apply the RC model, one needs to make two key assumptions. First, for each choice that is made by each node, one has to specify the set of alternatives and the time at which the choice occurs. Second, one needs to specify which observable characteristics influence the choices that are made by the nodes. Additionally, these observable characteristics must be deterministic when the choices occur.

We demonstrate how to apply the RC model to networks with edge independence and `sequential networks' (i.e., networks in which edges form in a sequence).
As we will discuss (see {Section \ref{apply_rc}}), the RC model --- with appropriate assumptions --- is also relevant to a variety of network modeling frameworks, such as stochastic actor-oriented models (SAOMs) \cite{snijders2002markov} and latent-space models (LSMs) \cite{hoff2002latent}. 
However, there may be unanticipated challenges to applying the RC model to such frameworks, and this has the potential to limit the applicability of the RC model.
Exploring the applicability and appropriateness of the RC model to different modeling frameworks is a worthwhile avenue for future work.

The present paper proceeds as follows. In Section \ref{LitReview}, we survey the existing literature on network formation and discuss our contributions in the context of this prior work. In Section \ref{sec:Discrete-Choice-Models},
we overview DC models. In Section \ref{sec:Mixed-Logit-Models},
we discuss MX models and the RC model. 
In Section \ref{sec:Synthetic-Network}, we apply the RC model to a set of synthetic networks. In Section \ref{sec:Citation Network}, we apply it to a patent citation network.
In Section \ref{sec:Limitations-and-Applicability}, we discuss the
limitations and applicability of the RC model. We summarize and discuss our main results in Section \ref{sec:Conclusions-and-Discussions}. In Appendix \ref{appendix1}, we give some additional details of our comparison between the RC and MNL models.


\section{Literature Review}\label{LitReview}

There is a vast literature on network formation that crosses many disciplines \cite{albert2002statistical,goldenberg2010survey,blume2011identification,graham2015methods,chandrasekhar2016econometrics,newman2010networks, cranmer2020inferential}. 
In this section, we discuss how our paper relates to the existing literature.

In economics, there are several models of strategic network
formation. See \cite{jackson2005survey} and chapters 5, 6, and 11 of \cite{jackson2010social} for surveys of such models. 
The basic idea behind most of these models is that each node chooses which edges to form to other nodes based on some `utility' that the node obtains from the resulting network. An important feature of these models 
is that nodes have at least partial control over which edges they form. 
Inspired by previous studies of strategic network formation \cite{blochjackson2006definitions}, our paper examines
edge formation that results from choices that are made by the nodes of a network. We explore this conceptual idea using DC models.

Several studies have employed DC models to study network formation. For example, Overgoor \emph{et al.} \cite{DiscreteChoice} used DC models to examine sequential networks and Yeung \cite{yeung2019statistical} used DC models to estimate the preferences of participants in bipartite networks. Other researchers have used DC models as building blocks in network-formation models~\cite{christakis2010empirical,goldsmith2013social,graham2014a,konig2016formation, tomlinson2022graph}. In SAOMs, which are a popular type of network-formation models, actors make myopic\footnote{A `myopic' choice is a choice that depends only on the current state of a network; it does not take into account changes in the network.} and binary changes to a network at discrete times~\cite{snijders2001statistical, snijders2010introduction}. 
Such changes are often studied using the MNL model. LSMs have employed the MNL model to study how positions in an unobserved space affect edge formation \cite{hoff2002latent}.

Most studies in the network-formation literature that have used DC models have focused on the MNL model. 
In Section \ref{subsec:LimitationsMNL}, we argue that the standard MNL model has several limitations when studying network formation. We then argue in Section \ref{subsec:Repeated-Choice-Model} that the RC model, which is also a DC model, has several desirable features that helps 
overcome the aforementioned limitations. In principle, other extensions of the MNL model may also overcome some of these limitations. For example, recent extensions of the MNL model allow the set of alternatives to influence an individual's preferences \cite{tomlinson2020learning,tomlinson2021choice}.
However, to the best of our knowledge, these extensions have not been applied to study network formation in a way that overcomes the limitations that we highlight.

An important difference between the RC and MNL models is that the parameters of the RC model are unbiased when node preferences are either heterogeneous or homogeneous, whereas the parameters of the MNL model are unbiased only when the preferences are homogeneous. We illustrate this difference using 
a set of synthetic networks. Graham \cite{graham2014a} studied heterogeneity in network formation using fixed effects.\footnote{In the context of network formation, `fixed effects' refer to an individual's specific propensity to form an edge that does not change with time.
See Wooldridge \cite{wooldridge2016introductory} for a detailed treatment of fixed effects.} We use the RC model to study heterogeneity in edge formation that may arise due to heterogeneity in preferences.

There is a class of network-formation models, which are sometimes called `edge-independent models' \cite{chandrasekhar2016econometrics}, that make the assumption that each edge in a network forms independently of all other edges. Examples of edge-independent network models include Erd\H{o}s--R\'{e}nyi (ER) graphs \cite{newman2010networks}, random geometric graphs (RGGs) \cite{penrose2003random}, and risk-sharing models \cite{fafchamps2007risk}. Using a set of synthetic networks, we demonstrate how to use the RC model to estimate parameters in edge-independent models of networks. We also illustrate how to use the RC model to estimate parameters in preferential-attachment (PA) network models \cite{Yule1925,Simon1955,Price76ageneral,newman2010networks}. In a PA model, one begins with a seed network (which typically is small) and new nodes arrive and connect to the network as time marches on. Each new node connects to one or more existing nodes with a probability that is determined by a specified kernel. Estimating parameters in a PA model can help improve understanding of how a node chooses which edges to form.  As an example of PA, we study a patent citation network using our RC model. 
 
 Other models have also been used to study sequential network formation.
 Recently, exponential random-graph models (ERGMs) were used to study large citation networks with similar assumptions as ours \cite{schmid2021generative, chakraborty2020patent}. In Section \ref{subsec:cite-comparison-models}, we compare the RC model and ERGMs in the context of citation networks. 
  Another relevant model is the latent ordered logistic (LOLOG) model \cite{fellows2018new}, which considers sequential network formation in a way that is similar to PA models.
  We also highlight relational-event models (REMs), which are particular popular in studies of sequential network formation \cite{butts20084}. REMs have been used to explore a variety of networks, such as communication networks during the World Trade Center disaster \cite{butts20084}, collaborations in the United States Congress \cite{brandenberger2018trading}, and corporate networks \cite{valeeva2020duality}. To the best of our knowledge, the parameters of the REM models in the current literature are homogeneous across the nodes of a network \cite{butts20084}, whereas the parameters of the RC model can be heterogeneous across the nodes of a network.
 

\section{Discrete-Choice (DC) Models}
\label{sec:Discrete-Choice-Models}

A discrete choice is an event in which an agent chooses between two or more discrete alternatives. 
For example, a person who is buying a new car is making a discrete choice \cite{train1986qualitative}.
In economics and statistics, it is common to use DC models to analyze such decisions \cite{kleinbaum2002logistic,HosmerLemesbow,menard2002applied,Train_Discrete_Book}. In a DC model, a node $n$ chooses from $J = |C|$ alternatives, where $C$ denotes the set of alternatives. For concreteness, suppose that each node represents an individual.
 Each alternative $j$ has a set of observable characteristics $x_{nj}$ that influence individual $n$'s decision.\footnote{It is common in DC models to assume that the set of alternatives does not directly influence the choice decisions \cite{train1986qualitative}. Tomlinson and Benson \cite{tomlinson2020learning} recently relaxed this assumption in the context of network formation. They allowed the context of the set of alternatives to influence a chooser's decision. For example, the decision to order a hamburger or pizza may be influenced by whether or not french fries are also on a menu.}


{\subsection{DC Models and Network-Formation Models}

One can view network formation as a consequence of `choices'
that are made by nodes. Each node `chooses' to form edges (i.e., `ties') to other nodes.
For example, in a directed network, if there is an edge from node $i$
to node $j$, then node $i$ `chose' to form an edge to node
$j$. Each node has certain characteristics that determines the attachment
probability for a node to select it; such choices determine how a network grows.
The assumption that edges form because of the choices of nodes is part of several network-formation models \cite{snijders2001statistical, snijders2010introduction, mele2017structural, fellows2018new, chandrasekhar2016econometrics}. In some real-world networks, this assumption is literally true because the nodes represent agents who are making choices. For example, individuals typically choose their friendships in social networks,
authors choose which papers\footnote{Naturally, referees also play a role in this process.} and patents to cite in citation networks, and government agencies choose where to build roads in transportation networks. In many other networks, nodes are not literally making choices, but one can still use a DC model if observable characteristics influence which edges form in the network. Although the assumption that edges form because of the choices of nodes is reasonable in investigations of many real-world networks, it is not appropriate for all networks. For example, this assumption may not be appropriate when studying a protein--protein interaction network because the edges in such a network form because of biochemical processes, rather than because of choices that are made by proteins.

It is convenient to divide a network-formation model into two components.
The \emph{network process} describes how a choice situation
arises. It primarily concerns when nodes form attachments, the set of available choices, and the set of characteristics that affect choices. 
The \emph{choice} component entails how nodes make
choices  when a choice situation arises.

Researchers use DC models to study how a node $n$ forms attachments
\emph{conditional} on the set $C$ of available choices 
and the deterministic observable characteristics $x_{nj}$ of each choice $j$ \cite{Train_Discrete_Book}.
A DC model is a choice component of a network-formation model. 
Many network-formation models employ a DC model as their choice component. For example, in an SAOM, actors have the opportunity to change one edge 
at random times that are based on the state of the network at that 
time \cite{snijders2010introduction}. These features
constitute the network-process component of such a network-formation model.
In the choice component of an SAOM, {it is common to use the} MNL to study how nodes change ties when given the opportunity to do so \cite{snijders2010introduction}. 

In some studies, the network-process component is known or observed because of characteristics of the focal application. For example, Overgoor \emph{et al.} \cite{DiscreteChoice} examined a social network from the social-media platform Flickr. 
In their data set, they possessed observations of exactly when each node made a change, what choices were available for those nodes, and the characteristics of the nodes that affect these choices\footnote{One example of a characteristic is the number of followers that a node has on Flickr.}.
 Consequently, they were able to directly apply the original MNL model to study network formation. 
 In other words, because they observed the network-process component, they did not need to model it; it was sufficient for them to model the choice component.  

Different papers have made different assumptions about the network-process component while using a DC model for the choice component.  
For example, in some edge-independent network models \cite{chandrasekhar2016econometrics}, it has been assumed that all edge-formation decisions occur simultaneously and independently. In contrast to such edge-independent models, the nodes in most SAOMs make edge-formation decisions one at a time (i.e., sequentially)
and at randomly determined points in time \cite{snijders2001statistical}. In many PA models \cite{newman2010networks}, nodes appear sequentially as a network grows and each node forms edges that attach to existing nodes only when it first appears. LSMs model the network-process component using positions in an unobserved social space\footnote{{A `social space' is a space of latent characteristics that affect the probability of edge formation in a network \cite{hoff2002latent}.}} and 
have used the MNL model for the choice component \cite{hoff2002latent}. 

In the present paper, we make no general claims about the network-process component of a network-formation model; instead, we focus our attention on the use of DC models in the choice component of such a model. However, for concreteness, we do make specific assumptions about the network-process component in our case studies. }


\subsection{Benefits of DC Models}\label{Advantages_DC}

DC models have several desirable properties for the study of growing networks. We briefly overview these benefits.

First, using tools such as logistic regression, one can apply DC models to networks that one constructs from empirical data; one can also readily estimate the parameters of such models~\cite{Train_Discrete_Book}. 

Second, DC models have been studied extensively in fields such as economics and statistics. These studies have led to tools for estimation, hypothesis
testing, and parameter selection \cite{HosmerLemesbow,kleinbaum2002logistic,menard2002applied,Train_Discrete_Book}. Additionally, it has been established that the estimators for many DC models (including the MNL and RC models that we examine) are unbiased and consistent in a statistical sense\footnote{An estimator of a parameter is `consistent' if it converges in probability to the true value of the parameter. An estimator is `unbiased' if the expectation of the estimator is equal to the true value of the parameter.}~\cite{HosmerLemesbow,kleinbaum2002logistic,menard2002applied,Train_Discrete_Book}.

Third, DC models are very flexible. Through appropriate choices of variables, one can use them to obtain several types of network-formation models. For example, Overgoor \emph{et al.} \cite{DiscreteChoice} showed that one can use the original MNL model to construct preferential-attachment (PA) models (such as the ones in~\cite{Bianconi_2001,Price76ageneral}), the $G(N,p)$ ER model \cite{erdos59a}, and triadic-closure models~\cite{DiscreteChoice}. 
One can also use DC models to construct a network-formation model that combines features of two or more existing models. Moreover, given a network, a DC model allows one to compare the relative importances of different network features {in the network-formation process.}


\subsection{The Basic Structure of DC Models}
\label{subsec:Structure}

In a DC model, an individual $n$ chooses from $J = |C|$ alternatives, where $C$ denotes the set of alternatives. Each alternative $j$ provides $n$ with a utility that depends on certain observable characteristics $x_{nj}$. In principle, the effect
of the observable characteristics can take any functional form. However,
almost all studies have assumed that the relationship is linear \cite{Train_Discrete_Book}.
Following this convention, we define the \emph{observable utility}
to be $U_{nj}=\beta^{T}x_{nj}$ \cite{Train_Discrete_Book}. The parameters $\beta$ are {called} \emph{choice parameters} and reflect the \emph{preferences}\footnote{The preferences of an individual $j$ capture the amount that the characteristic $x_{nj}$ affects the utility of a choice by $j$.}
of an individual $j$. The utility also has a random component $\epsilon_{nj}$. 
The \emph{total utility} of individual $n$ of making choice $j$ is $V_{nj}=U_{nj}+\epsilon_{nj}$.
Individual $n$ chooses an alternative $j$ if it provides them with
more utility than any other alternative.\footnote{The probability that any two alternatives have the same utility is $0$ because the random component is a continuous
random variable.} It is often assumed that $\epsilon_{nj}$ follows a probability distribution that is an independently and identically distributed (IID) standard type-1 extreme value distribution\footnote{The cumulative distribution function of the standard type-1 extreme-value distribution is $F(x) = \exp\{-e^{-x}\}$.} (i.e., it is a Gumbel distribution) \cite{Train_Discrete_Book}.

Assuming that $U_{nj}=\beta^{T}x_{nj}$ and that $\epsilon_{nj}$ follows an IID type-1 extreme-value distribution, the probability that individual $n$ chooses alternative
$j$ is
\begin{equation}\label{eq:prob}
	P_{nj}=\frac{\exp(\beta^{T}x_{nj})}{\sum_{i\in C}\exp(\beta^{T}x_{ni})}\,.
\end{equation}
This particular formulation is the MNL model\footnote{{This formulation is also known as the `conditional MNL model'. Following \cite{Train_Discrete_Book}, we refer to it as simply the `MNL model'.}}~\cite{Train_Discrete_Book}. 

One can readily estimate the choice parameter $\beta$ 
using maximum-likelihood estimation (MLE), and there exist
several methods for this statistical inference \cite{Train_Discrete_Book}. The MNL model is one of the most common
DC models because it is statistically simple and is supported by a well-developed theory.
For a thorough treatment of the MNL model, see Kleinbaum and Klein \cite{kleinbaum2002logistic} or Train \cite{Train_Discrete_Book}.

The MNL model requires certain assumptions that are often violated when studying network formation.
We discuss these assumptions and associated limitations in Section \ref{subsec:LimitationsMNL}. 


\subsection{Limitations of the Multinomial Logit (MNL) Model \label{subsec:LimitationsMNL}}

The MNL model has two key limitations when studying network formation.\footnote{Another limitation of the MNL model is that it assumes independence
of irrelevant alternatives (IIA). The principle of IIA states that when choosing between choices $x$ and $y$, other alternatives do not affect one's choice. 
That is, the ratio of the probability of choice $x$ to the probability of choice $y$ does not depend on other alternatives, where the probability of a choice is defined by equation \eqref{eq:prob}. The assumption of IIA is contested hotly in the decision-theory literature \cite{hanks_just_wansink_2012,BensonIIAViolation}. When using the MNL model to study network formation, that model inherits the IIA assumption. As was shown in Hanks \emph{et al.} \cite{hanks_just_wansink_2012} and Benson \emph{et al.}~\cite{BensonIIAViolation}, this assumption is often violated. See page 54 of \cite{Train_Discrete_Book} for a simple example in which IIA fails. To apply the MNL model to study network formation, it is important to check the validity of the IIA assumption. If it is violated, then it is likely undesirable to use the MNL model because one of its central assumptions is violated. Other types of models, such as probit models, do not require the IIA assumption. However, probit models are both more expensive computationally than logit models and are more cumbersome to extend than logit models.} These limitations restrict the types of networks for which the MNL model is appropriate. Because these limitations arise from the structure of the MNL model, it is necessary either to generalize the MNL model or to use another type of model to overcome them.

The first key limitation is that the MNL model assumes that a node chooses only one of the $J$ alternatives. 
This implies that the out-degree of any new node is equal to $1$ when it is added to a network. However, in most situations,
this is rarely true in practice~\cite{newman2010networks}. 
This requirement may lead to biases in estimates from using the MNL model in studies of network formation. These biases arise because observations may no longer be IID; there may be correlations between the choices of a given node at different times.
For example, in a social network, an individual's choices of friends at different times
are likely to be correlated with each other \cite{socialanalysisbook}. Similarly, we do not expect a paper to cite other papers in an independent way \cite{slyder2011citation}.

The second key limitation is that the MNL model assumes implicitly that each individual (i.e.,
each node) has the same choice set. In the context of network formation,
this assumption is often inaccurate. If we consider sequential network formation, in which new nodes appear one after another,
 newer nodes typically have a larger choice set than older nodes. For example, newer papers necessarily have a larger set of papers to cite
than their predecessors. Other features of a network can also result in different nodes having different choice sets. The presence
of heterogeneous choice sets also leads to biases in the estimates of the
MNL model \cite{Train_Discrete_Book}.

These limitations are not applicable to all network-formation models. For example, many SAOMs assume that the number of nodes is fixed and that all edge-formation decisions are independent \cite{snijders2017stochastic}. Under these assumptions, the two key limitations of the MNL model are irrelevant.


\section{Mixed Logit (MX) Models}
\label{sec:Mixed-Logit-Models}

Mixed logit (MX) models are flexible generalizations of the MNL model \cite{Train_Discrete_Book}. MX models can approximate
any random-utility model\footnote{In a random-utility model, the utility (i.e., benefit)
of a choosing an alternative is a function of one or more random variables.}~\cite{McFaTrai00}. 
In an MX model, the choice parameters $\beta$ can be heterogeneous across individuals.
We use this feature of MX models to overcome the associated limitation of the MNL model that we described in Section \ref{subsec:LimitationsMNL}.

Let $\beta_{n}$ denote the choice parameter for individual $n$. For example, in a citation network, the effect of past citations can be different
for different papers. Hypothetically, some discipline $u$ may be more likely to cite more papers than another discipline $v$; one can capture this difference with the inequality $\beta_{u}>\beta_{v}$. One captures the heterogeneity of the choice parameters across individuals by letting
\begin{equation}
	V_{nj}=\beta_{n}^{T}x_{nj}+\epsilon_{nj}\,.
\end{equation}

We draw the choice parameters from a 
distribution $f(\beta_{n}|\theta)$,
where $\theta$ is a vector that encodes the distribution parameters.
The distribution parameters specify the distribution of the choice parameters $\beta_{n}$. For example, if $f(\beta_{n}|\theta)$
is normally distributed, then $\theta$ consists of
the mean and the variance.

For a specified value of $\beta_{n}$, we write the conditional choice probability as
\begin{equation}
    \label{mx_probability}
	P_{nj}(\beta_{n})=\frac{\exp(\beta_{n}^{T}x_{nj})}{\sum_{i\in C_{n}}\exp(\beta_{n}^{T}x_{ni})}\,,
\end{equation}
where $C_{n}$ is individual $n$'s choice set. The choice set can be different for different individuals.\footnote{Yamamoto~\cite{Yamamoto2010AMR} studied a special case of MX models for situations in which the choice set is different for different groups of individuals.} 
To compute the unconditional probability, we take the expectation of the probability measure of
$f(\beta_{n}|\theta)$ by calculating 
\begin{equation}
	P_{nj}=\int_{\beta_{n}\in D_{\beta_{n}}}P_{nj}(\beta_{n})f(\beta_{n}|\theta) \, \mathrm{d}\beta_{n}\,,
\end{equation}
where $D_{\beta_{n}}$ is the support of $f(\beta_{n}|\theta)$. We use MLE and unconditional probabilities to estimate $\theta$.

In Section \ref{subsec:LimitationsMNL}, we highlighted the fact that the MNL
model assumes that each node chooses only a single alternative. 
In the context of network formation, this assumption implies that all new nodes have an out-degree of $1$, which is almost never true in practice~\cite{newman2010networks}.
However, it is challenging to use MX models to allow multiple choices. From
a theoretical perspective, one can expand the choice set to include
multiple choices. If one does so, to have a valid probability measure,
one must also expand the choice set to be the power set of all of the single-choice
alternatives. This results in a choice set with $2^{|C|}$ elements (recall that $C$ denotes the choice set), which can 
render it computationally intractable to perform estimation. Furthermore, from an intuitive perspective, it is not clear how to define and interpret the utility of multiple selections in a network context. For the process of selecting one item $\{j\}$, one can define the utility as $U_{nj}=\beta_{n}^{T}x_{nj}$.
However, it is not clear how to define the utility of selecting two items $\{j,k\}$. To resolve this issue, we view multiple selections as a sequence of choices. 
Considering multiple selections as a choice sequence allows us to estimate the distribution of $\beta_{n}$ using the RC model \cite{TrainRevelt97mixedlogit}, which is a special 
type of MX model. 


\subsection{The Repeated-Choice (RC) Model}
\label{subsec:Repeated-Choice-Model}

The RC model, which was first proposed in~\cite{TrainRevelt97mixedlogit}, is well-suited to
studying network formation. The RC model allows one to study networks with nodes whose out-degree
is larger than $1$. Additionally, the out-degree and choice set can be different
for different nodes. 

In the RC model, an individual $n$ faces a situation
in which they choose from $J$ alternatives at each of $T_{n}$ times.
The number of times at which individual $n$ makes a choice can be different for different individuals, and the set of alternatives can also be 
 heterogeneous both across individuals and across 
choice instances of the same individual. Similarly to Equation \eqref{mx_probability}, the attachment probability is 
\begin{equation} \label{eq:choice_prob}
	P_{njt}(\beta_{n})=\frac{\exp(\beta_{n}^{T}x_{njt})}{\sum_{i\in C_{nt}}\exp(\beta_{n}^{T}x_{nit})}\,,
\end{equation}
where $P_{njt}$ is the probability that individual $n$ chooses alternative
$j$ at time $t$. The vector $x_{njt}$ encodes the covariates
for individual $n$ choosing alternative $j$ at discrete time $t$, and $C_{nt}$
denotes the set of alternatives that are available to individual $n$ at time $t$.

Let $i(n,t)$ denote the choice that is made by individual $n$ at time $t$. 
Because we are examining repeated choices across time, we consider the sequence $\{i(n,1),i(n,2),\ldots, i(n,T_{n})\}$ of
choices of individual $n$. Given a value of $\beta_{n}$, these choices are independent. 
Therefore, we can write the conditional probability of a sequence of choices as
\begin{equation}
	S_{n}(\beta_{n})=\prod_{t=1}^{T_{n}}P_{ni(n,t)t}(\beta_{n})\,.
\end{equation}
Taking an expectation over $\beta_{n}$ yields the unconditional probability
\begin{equation} \label{eq:expected_prob}
	P_{n}(\theta)=\int_{\beta_{n}\in D_{\beta_{n}}}S_{n}(\beta_{n})f(\beta_{n}|\theta) \, \mathrm{d}\beta_{n}\,,
\end{equation}
where $D_{\beta_{n}}$ is the support of $f(\beta_{n}|\theta)$.

We use the unconditional probability $P_{n}(\theta)$ to estimate $\theta$, which determines the distribution of $\beta_{n}$. If $\beta_{n}$ is high-dimensional,\footnote{{Recall that $\beta_{n}\in\mathbb{R}^{k}$ is a vector of parameters. We refer to $\beta_{n}$ as `high-dimensional' if $k$ is large, although different disciplines have different notions of what it means to be `large'. In statistics, $k$ is often interpreted as large if $k\approx N$, where $N$ is the number of observations \cite{buhlmann2011statistics}.}} it is difficult to analytically compute $P_{n}(\theta)$, so we use maximum-simulated-likelihood estimation (MSLE) to determine consistent estimators of the distribution parameters.
The simulated log-likelihood function is $L(\theta)=\sum_{n}\ln[P_{n}(\theta)]$, and we use simulations to calculate $P_{n}(\theta)$. For a given value of $\theta$, we draw several values of $\beta_n$ from the distribution $f(\beta_{n}|\theta)$. In our case studies, we draw values of $\beta_n$ $100$ times from $f(\beta_{n}|\theta)$; a good choice for the number of draws depends both on the situation and on the availability of computational resources. For each draw, we compute $S_{n}(\beta_{n})$. The mean of $S_{n}(\beta_{n})$ across all draws is an asymptotically consistent estimator of $P_{n}(\theta)$. See \cite{Train_Discrete_Book,TrainRevelt97mixedlogit} for detailed discussions of simulation and estimation methods. As discussed in \cite{TrainRevelt97mixedlogit,Train_Discrete_Book}, the parameter estimates are also unbiased.


\subsection{Applying the RC Model to Network Data}
\label{apply_rc}

To use the RC model to study network formation, we need to translate network data
into an ordered set of \emph{choice sequences}. 
As the name suggests, a choice sequence is a collection
of \emph{choices}.\footnote{Depending on the application and the question of interest, the order of the choices in a choice sequence may or may not be important.} There is one choice sequence for each node in a network whose out-degree is at least $1$. 
The number of choices\footnote{In the economics literature, the number of choices is sometimes called the number of `time periods' \cite{TrainRevelt97mixedlogit}. To avoid confusion, we do not adopt this convention in our paper.} in a sequence is equal to the out-degree of the corresponding node.
Each choice consists of a node and a set of alternatives. 

As an example, consider a network with nodes $\{p,q,r,s\}$ and directed
edges $\{(p,q),(p,r),(q,s),(r,s)\}$. We view the network as a set
of three choice sequences because there are three nodes ($p$, $q$, and $r$) whose out-degree is at least $1$. We denote the choice sequences
for nodes $p$, $q$, and $r$ by $S_{p}$, $S_{q}$, and $S_{r}$, respectively. 

The sequence $S_{p}$ consists of two choices because
the out-degree of node $p$ is $2$. Specifically, $S_{p}=\{(i(p,1),C_{p_1}),(i(p,2),C_{p_2})\}$.
The tuple $(i(p,1),C_{p_1})$ represents the first `choice' that is made
by node $p$. The tuple's first element $i(p,1))$ indicates the first `chosen' node. 
The order of the choices in a choice sequence is inconsequential for estimation\footnote{A `choice sequence' is  technically a set. We refrain from referring to a `choice sequence' as a `choice set' because it is common to use the term `choice set' to refer to an `alternative set'.} \cite{Train_Discrete_Book}.
However, depending on the application and question of interest, one order may be more convenient than others for constructing the choice sequence.
 For example, if we consider a network that is changing with time, it may be convenient to order the choice sequence using chronological ordering.
We arbitrarily set $i(p,1)$ to be equal
to $q$ (rather than to $r)$. The tuple's second element $C_{p_1}$ indicates the set of alternatives (i.e., $C_{p_1}$) that are available to the node (i.e., node $p$) that
is `choosing'. 

For each choice, the set of alternatives includes all nodes except the node that is choosing. The alternative set $C_{p_1}$ is equal to $\{q,r,s\}$.
We do not include node $p$ in the alternative set because we
do not allow self-edges. By the same convention, $i(p,2)$ is equal to $r$ and
$C_{p_2}$ is equal to $\{q,r,s\}$. 
We use the same convention to construct
the choice sequences for nodes $q$ and $r$. Specifically,
$S_{q}=\{(i(q,1),C_{q_1})\}$ and $S_{r}=\{(i(r,1),C_{r_1})\}$, where
both $i(q,1$) and $i(r,1)$ are equal to $s$, the alternative set $C_{q_1}$ is equal
to $\{p,r,s\},$ and $C_{r_1}$ is equal to $\{p,q,s\}.$ We do not
construct a choice sequence for node $s$ because its out-degree is $0$. We can directly apply the RC model to the set of choice
sequences to estimate the distribution parameters.

In our setting, we need to make two key decisions to apply the RC model to study network formation. First, for each choice of each node, we have to decide which nodes are in the set of alternatives. In the simple example that we discussed in the preceding paragraphs, we assumed for a given node that all other nodes are in the set of alternatives. However, such an assumption is not appropriate in all applications. For example, if a network is generated by the standard RGG model \cite{penrose2003random}, all nodes within a certain radius of a given node are in the set of alternatives for all of the choices of that node (and no other nodes are in the set of alternatives). Additionally, the set of alternatives can be different for different choices of the same node. Second, we have to decide which observable characteristics (i.e., which $x_{njt}$) influence the choices of a node. For example, in some PA models \cite{Bianconi_2001}, the in-degree of existing nodes influences the choices of a new node. A key assumption in the RC model is that the observable characteristics are deterministic when a node makes a choice; this restricts the set of observable characteristics that one can study using the RC model. For example, the observable characteristics cannot include variables that are influenced by the choices of subsequent nodes because (by construction) the choices of future nodes are stochastic.\footnote{This assumption does not require that a node simultaneously form all of its edges to a network. A node can form edges at different times, and other nodes can form edges in time intervals between two choices of a given node. The only requirement is that any observable characteristics that influence edge-formation are deterministic when each node makes each choice.}
In Section \ref{sec:Citation Network}, we use a case study of patent citation networks to illustrate which observable characteristics are reasonable to treat as deterministic when a node makes a choice. {In certain situations, such as in sequential networks and in edge-independent networks\footnote{For example, if one observes one snapshot of a network, it is common to assume edge independence.}, the assumption that observational characteristics are deterministic is equivalent to assuming that the observable characteristics are sufficient statistics. In these situations, conditional on these statistics, edge-formation choices are independent.\footnote{{However, the assumption that observable characteristics are deterministic and the assumption that the observable characteristics are sufficient statistics are not equivalent in all situations. {For example, {in an SAOM,} the observable characteristics are not sufficient statistics \cite{snijders2010introduction}.}}}} 

The above decisions depend critically on the network-process component of a network-formation model. 
In our simple example, we only observed one snapshot of a network. Without additional assumptions or data, we cannot include the edge between $q$ and $r$ as an observable characteristic when using the RC model to study the decision of node $p$ to link to node $q$ because the decision of node $q$ to link to node $r$ is stochastic. 
This situation illustrates the so-called `endogeneity problem'.  

The assumption that observable characteristics are deterministic does not imply that one cannot use the RC model to study
such features or that the RC model requires strong assumptions like edge independence. 
If one possesses additional data about the timing of edge formation, then one can use the formation of edges in the past
as observable characteristics because the past decisions that resulted in those edges are deterministic. 

In our case study of a citation network in Section \ref{sec:Citation Network}, we use the RC model to examine whether or not new patents are more likely to cite patents with 
larger citation counts. We are able to do so because of the sequential edge formation in the patent citation network. We sometimes refer to such a network itself as `sequential'. We argue why the assumption of sequential edge formation is appropriate in the context of our application to a patent citation network. However, this assumption is not necessary when using the RC model to study whether or not previous edge-formation decisions influence future edge-formation decisions.
For example, one can potentially use the RC model in conjunction with an SAOM model without assuming that a network is sequential. Employing the RC model in this way an avenue for future research.

We also need to assume that the choice parameters satisfy a distribution of a specified type.
In general, it is difficult to know what type of distribution is appropriate for a particular application, and one usually cannot evaluate such an assumption. In our case studies, we make our assumptions about distributions based on computational convenience and prior knowledge about our applications of interest. 


\section{Case Study: A Set of Synthetic Networks}
\label{sec:Synthetic-Network}

We illustrate the RC model using a set of synthetic networks, which are useful to study for several reasons. 
First, we know the true data-generation process of such networks, so we can use them to directly evaluate the performance of our model. Second, we control the sizes of each set of synthetic networks and of each network in the set. Third, we can choose a process that allows us to generate data that is free from data-collection errors, which arise often in networks that one constructs from empirical data. 

We generate the network data using the RC model. In Section \ref{subsec:SDatanadModel}, we describe our data-generation procedure and model specification. 
If the RC model does not perform well for synthetic data that it generates, then it is unlikely to be successful for networks that one constructs from empirical data.


\subsection{Data Construction and Model Specification}
\label{subsec:SDatanadModel}

We construct a set of $100$ synthetic networks in which each network has $1000$ nodes. We draw the out-degree of each node uniformly at random and independently from $\{1, \ldots, 20\}$. As we described in Section \ref{subsec:Repeated-Choice-Model}, each node has an associated choice sequence. The length of a node's choice sequence is equal to its out-degree. {Each choice in a choice sequence consists of (1) a `chosen node' that is selected by the focal node and (2) all other nodes that the focal node could have chosen instead (i.e., the set of alternative nodes).} For each set of alternatives, we draw the number of alternatives uniformly at random from $\{1, \ldots, 10\}$. For each choice, the probability that an alternative is chosen depends on two quantities. Specifically, the probability that node $n$ chooses alternative $j$ as its $t^{\textrm{th}}$
choice depends on $x_{njt}$ and $y_{njt}$. The subscript
$njt$ denotes the $j^{\textrm{th}}$ alternative that is available to node
$n$ for the $t^{\textrm{th}}$ choice in {its} choice sequence. We sample $x_{njt}$ from a continuous uniform distribution between $-1$
and $1$ (i.e., $x_{njt}\sim\mathcal{U}[-1,1])$, and we sample $y_{njt}$
from a continuous uniform distribution between $0$ and $5$ (i.e.,
$y_{njt}\sim\mathcal{U}[0,5]$). Using the formulation that we described
in Section \ref{sec:Mixed-Logit-Models}, the choice utility is given by the function
\begin{equation} \label{eg:synth-utility}
	V_{njt}=\beta_{1n}x_{njt}+\beta_{2n}y_{njt}+\epsilon_{njt}\,,
\end{equation}
where $\beta_{1n}$ and $\beta_{2n}$ are the choice parameters and
we draw $\epsilon_{njt}$ from an IID Gumbel distribution. The distributions
of the choice parameters are $\beta_{1n}\sim\mathcal{N}(3,4)$ and $\beta_{2n}\sim\textrm{LogNormal}(0,1)$. With $Z\sim\mathcal{N}(a,b)$ and $X=\exp(Z)$, it follows that $X$ is distributed log-normally with parameters $a$ and $b$ (i.e., $X\sim\textrm{LogNormal}(a,b)$).

For each choice in each choice sequence, we draw the `chosen' alternative
using the probability mass function (PMF) in equation
\eqref{eq:choice_prob}. The chosen alternative and the alternative
set together constitute a choice. We repeat this process for each choice
in each of the choice sequences. In our model, recall that the set of choice
sequences encodes a network.

We apply the RC model to each network in our set of networks. 
To do this, we first need to know the distribution type of the
choice parameters. We refer to this assumption as the \emph{distributional
assumption}. We then use the RC model to estimate the distribution
parameters. In almost all situations, one does not know the true distribution type
of the choice parameters.

We do know the true distributions of the choice parameters in our synthetic networks because we generated these networks. Specifically, we know that the choice parameters satisfy $\beta_{1n}\sim\mathcal{N}(m_{1},s_{1}^{2})$ and $\beta_{2n}\sim\textrm{LogNormal}(m_{2},s_{2}^{2})$. We use our model to estimate the distribution parameters $m_{1}$, $s_{1}$, $m_{2}$, and $s_{2}$. 

For networks that one constructs from empirical data, one does not know the true distributions, so one guesses 
the distribution type(s) of the choice parameters. (The requirement that one knows the distribution type of the 
choice parameters is a requirement of the RC model.) This is a difficult task; in general, one cannot even evaluate the distributional assumption, although there is some hope.
In Section \ref{subsec:SynMNLcomparison}, we demonstrate how to use the RC model to check whether the choice parameters are random
or deterministic. Unfortunately, if these parameters are random, we do not know how to evaluate whether or not a conjectured distribution type is the same as the true  distribution. This is a limitation of {the RC} model; we discuss this limitation in further
detail in Section \ref{sec:Limitations-and-Applicability}. From a practical perspective, making useful guesses about distribution types benefits from prior knowledge of an application. For an illustration, see Section \ref{subsec:CiteModel-Specification}.


\subsection{Estimation Procedure and Computational Complexity}
\label{subsec:SynEstimatiom}

As we described in Section \ref{subsec:Repeated-Choice-Model}, we
can use MSLE to {compute consistent estimates of} our distribution
parameters. We independently compute estimates of the distribution
parameters for each network in a set, and we use the {\sc mlogit}
package in {\tt R} for estimation \cite{Croissant_estimationof}. We use simulations (with 100 draws) to determine the unconditional probability of each choice sequence of probabilities from \eqref{eq:expected_prob}. We then use the Berndt--Hall--Hall--Hausman (BHHH) optimization algorithm to determine the distribution parameters that maximize the simulated log-likelihood \cite{Croissant_estimationof}. For more details, see \cite{Croissant_estimationof}. 

The combined estimation procedure for all of the networks in our set of synthetic networks took
approximately 18 hours on a computer with an i7 processor and 16 gigabytes
(GB) of random-access-memory (RAM). For our case study of a patent citation network, we use the same estimation procedure on the same computer.


\subsection{Results}
\label{subsec:SynResults}

We use our model to independently estimate $m_{1}$, $s_{1}$, $m_{2}$, and
$s_{2}$ for each network in our set of synthetic networks. This yields 
a set of estimates for each distribution parameter. We show summary
statistics of these estimates in Table \ref{tab:ResultsSyn}.

\begin{table}[H]
\caption{\textbf{Summary statistics of the estimated values of the distribution parameters for our synthetic networks.} 
The first column specifies the distribution parameters. The second column specifies the associated characteristics.
 For example, the parameter $m_{1}$ is the mean of the distribution of $\beta_{1n}$. As defined in Equation \eqref{eg:synth-utility}, the amount that $x_{njt}$ affects the choice utility depends on $\beta_{1n}$. The third column gives the true values of the parameters.
The fourth column gives the mean estimated values of the parameters.
The fifth column gives the median estimated values of the parameters.
The sixth column gives the biases of the estimated values of the parameters, where the bias is the {absolute value of the difference between the true value of a parameter and} its mean estimated value. The last column gives the standard errors of the estimated values of the parameters.
}
\label{tab:ResultsSyn}

\vspace{1cm}

\centering{}%
\begin{tabular}{p{1.5cm}  p{1.5cm}  p{1cm}  p{1.5cm}  p{1.5cm}  p{1.5cm} p{1.5cm}}
\hline 
Parameter & Characteristic & True Value & Mean & Median  & Bias & Standard Error\tabularnewline
\hline 
\hline 
$m_{1}$ & $x_{njt}$ & $3$ & 2.975   & $2.977$   & 0.025    &  $0.100$\tabularnewline
\hline 
$s_{1}$ & $x_{njt}$ & 2   & 2.035   & $2.033$   & 0.035    & $0.042$\tabularnewline
\hline 
$m_{2}$ & $y_{njt}$ & 0   & $0.002$ & $-0.004$  & 0.002    & $0.093$\tabularnewline
\hline 
$s_{2}$ & $y_{njt}$ & 1   & $1.004$ & $1.002$   & 0.004    & $0.050$\tabularnewline
\hline 
\end{tabular}
\end{table}

The mean and median estimated values of the parameters
are close to the true values of the parameters. In expectation, the difference
between the true and estimated values is less than $0.07$ for all
of the parameters. Additionally, the standard error is at most $0.100$ for all of the parameters. 
The bias, which is the {absolute value of the difference between the true value and the mean estimated values}\footnote{{For example, the bias {in} the first row is {$|3-\hat{m}_{1}|$}, where $3$ is the true value of the parameter $m_{1}$ and {$\hat{m}_{1} = 2.975$} is the mean of the estimated values.}}, is $0.025$, $0.035$, $0.002$, and $0.004$ for $m_{1}$, $s_{1}$, $m_{2}$,
and $s_{2}$, respectively.

Based on our results, we conclude that the RC model is able to accurately estimate the distribution
parameters for most of the networks in our set of synthetic networks.


\subsection{Comparison of the RC and MNL Models}
\label{subsec:SynMNLcomparison}

We now compare the performance of the RC model with that of the MNL model. We use a set of 50 synthetic networks, which we generate using the RC model. Each network has 100 nodes. We draw one network uniformly at random
from the set, and we present results from that
network for our comparison. In Appendix \ref{sec:AppendixRandomParameters}, we show summary statistics of
the results from all of the networks in the set. The model specification
and network construction are the same as in Section \ref{subsec:SDatanadModel}. 


\subsubsection{Random Choice Parameters}
\label{subsec:Random-Choice-Parameters}

The data-generation process that we described in Section \ref{subsec:SDatanadModel}
violates the assumptions of the MNL model. In our data, the number of choices and
the alternative set are both heterogeneous across individuals. Furthermore, the choice parameters are distributed randomly; by contrast,
the MNL model assumes that the choice parameters are deterministic.
We ignore these violations and fit the data to the MNL model.
We then compare the point estimates of the choice parameters that we compute using
the MNL model to the distribution of the choice parameters that we compute
using the RC model (see Table \ref{tab:ResultsSyn}).

\begin{table}[H]
\caption{\textbf{{Point estimates from the MNL model for the synthetic networks that we described in Section \ref{subsec:SynResults}.}} {The first column specifies the parameters. The second column specifies the associated characteristics.
For example, as defined in Equation {\eqref{eg:synth-utility}}, the amount that $x_{njt}$ affects the choice utility depends on $\beta_{1n}$. The third column
gives the estimated values of the parameters, the fourth column gives
the standard errors of these estimates, and the last column gives
the p-values. For statistical-significance levels that are larger than the p-value, the
coefficient is significantly different from $0$.}
}
\label{tab:SynMNLEstimates}
 \vspace{1cm}

\centering{}%
\begin{tabular}{p{1.5cm} p{1.5cm} p{2.0cm}  p{2.5cm}  p{2.5cm} }
\hline 
Parameter & Characteristic & Coefficient & {Standard Error} & p-value\tabularnewline
\hline 
\hline 
$\beta_{1}$ & $x_{njt}$ & 1.9258 & 0.0094 & $< 0.0001$\tabularnewline
\hline 
$\beta_{2}$ & $y_{njt}$ & 0.8052 & 0.0038 & $< 0.0001$\tabularnewline
\hline 
\end{tabular}
\end{table}

The point estimates from the MNL model yield statistically significant
results. However, they do not accurately capture the data-generation
process. Tables \ref{tab:ResultsSyn} and \ref{tab:SynMNLEstimates}
illustrate that the point estimates from the MNL model are different
in magnitude from the central values (both the mean and the median) of the
true distributions. Moreover, the MNL model paints an incorrect picture of the data-generation process.
The point estimates imply that the choice parameters are homogeneous
across individuals. However, by construction, we know that this is not
the case. The MNL model ignores this heterogeneity across individuals.
For networks that one constructs from empirical data, this may have a significant impact on the
interpretation of results because it may yield misleading values for
the choice parameters. We revisit the issue of interpretation in Section
\ref{subsec:CiteComparisonMNL}. In that discussion, we use a network of patent citations that we construct from empirical data.

The above discussion helps highlight some of the salient differences
between the results from the MNL and RC models. Given our data-generation process (which violates several assumptions of the MNL model), it is not surprising that the MNL produces incorrect results. Because we used the RC model to construct the data, it is also not surprising that the RC model produces accurate results. In Section \ref{subsec:Deterministic-Choice-Parameters}, we compare the results of the MNL and RC models when the choice parameters
are deterministic.


\subsubsection{Deterministic Choice Parameters}
\label{subsec:Deterministic-Choice-Parameters}

We construct networks in which the choice parameters are deterministic.
The data-generation process and model specification are the same as
in Section \ref{subsec:SDatanadModel}. The only difference is that
the choice parameters are deterministic, with $\beta_{1n} = -1$ and
$\beta_{2n} = 3$. The network is suitable for the MNL model because these
parameters are deterministic. Additionally, our assumption in the RC model that the
parameters are randomly distributed (specifically, $\beta\sim\mathcal{N}(m_{1},s_{1}^{2})$
and $\beta_{2n}\sim\textrm{LogNormal}(m_{2},s_{2}^{2})$) is incorrect. As in our computations in Section \ref{subsec:Random-Choice-Parameters},
we fit the misspecified\footnote{Intuitively, a parameter is misspecified if the assumed distribution of the parameter in a model differs from 
the parameter's true distribution. In our application, the RC model is misspecified because we assume that its parameters are random when they are deterministic.
See \cite{misspecification} for a formal definition of `misspecification'.} RC model to the
data and compute estimates for the distribution parameters. We report
the results from the MNL model and the misspecified RC model in Table
\ref{tab:SynDetermin}.

\begin{table}[H]
\caption{\textbf{{Point estimates from the MNL model (with parameters $\beta_{1n}$ and $\beta_{2n}$)
and the RC model (with parameters $m_{1}$, $m_{2}$, $s_{1}$, and $s_{2}$) for the synthetic networks that we generate using the MNL model.}}
{The first column specifies the parameters. The second column specifies the associated characteristics. For example, the parameter $m_{1}$ is the mean of the distribution of $\beta_{1n}$. As defined in Equation \eqref{eg:synth-utility}, the amount that $x_{njt}$ affects the choice utility depends on $\beta_{1n}$.} The third column gives the estimated
values of the parameters, the fourth column gives the standard errors
of these estimates, and the last column gives the p-values. For statistical-significance levels
that are larger than the p-value, the coefficient is significantly
different from $0$.
}
 \label{tab:SynDetermin}
 \vspace{1cm}

\begin{centering}
\begin{tabular}{p{1.5cm}  p{1.5cm} p{1.5cm}  p{2.5cm}  p{2.5cm} }
\hline 
Parameter & Characteristic & Coefficient & {Standard Error} & p-value\tabularnewline
\hline 
\hline 
$\beta_{1}$ & $x_{njt}$ & $-0.988$ & $0.035$ & $<0.0001^{***}$\tabularnewline
\hline 
$\beta_{2}$ & $y_{njt}$ & $2.963$ & $0.045$ & $<0.0001^{***}$\tabularnewline
\hline 
$m_{1}$ & $x_{njt}$ & $-0.807$ & $0.108$ & $<0.0001^{***}$\tabularnewline
\hline 
$s_{1}$ & $x_{njt}$ & $-0.054$ & $0.321$ & $0.866$\tabularnewline
\hline 
$m_{2}$ & $y_{njt}$ & $1.015$ & $0.051$ & $<0.0001^{***}$\tabularnewline
\hline 
$s_{2}$ & $y_{njt}$ & $0.084$ & $0.139$ & $0.546$\tabularnewline
\hline 
\end{tabular}\vspace{0.5cm}
\par\end{centering}
\textquoteleft {***}\textquoteright, \textquoteleft {**}\textquoteright ,
\textquoteleft {*}\textquoteright, and \textquoteleft {$\bullet$}'
represent statistical-significance levels of {less than 0.01\%}, precisely 0.1\%, 1\%, and 5\%, respectively.
The absence of asterisks or \textquoteleft {$\bullet$}'
indicates that the parameters are not significantly different from $0$.
\end{table}

As we expected, the MNL model accurately estimates the choice parameters.
However, although the RC model is misspecified for this network, the distribution
of the choice parameters are good approximations of the true deterministic
choice parameters. The mean of $\beta_{1n}$ (which is $-0.807$)
gives a reasonable ballpark estimate of the true value of $-1$, and the standard
deviation of $\beta_{1n}$ differs from $0$ with a probability that is less
than $0.134$. Similarly, the mean of $\beta_{2n}$ (which is
$2.770$) gives a reasonable ballpark estimate of the true value of $3$, and and the standard deviation of $\beta_{2n}$ differs from $0$ with a probability that is less than $0.454$. If a random variable has a standard deviation of $0$, then it is deterministic, with
a value that is equal to its mean. Our results imply that the standard deviation of the choice parameters is equal
to $0$ with a large probability; this suggests that the choice parameters
are deterministic with a large probability. In this case, the RC model
yields results that are reasonably accurate qualitatively,
despite the model's misspecification. Deterministic parameters are a special case of random parameters (because they are random parameters with a standard deviation of $0$). Consequently, the RC model can be helpful for providing accurate estimates for our synthetic networks even when the choice parameters are deterministic. By contrast, as we saw in Section \ref{subsec:Random-Choice-Parameters}, the MNL model may not provide accurate results when the choice parameters are random. Moreover, one should not expect it to provide accurate results in such situations.

The above exercise also highlights a way to use the RC model to test whether choice parameters are random or deterministic. We calculate estimates under the assumption that the choice parameters are random. If the standard deviation
of the choice parameters is equal to $0$ with a large probability, then
it is likely that the choice parameters are deterministic. This is an advantage of the RC model over the MNL model. The estimates
of the MNL model provide no way to check whether parameters are
random or deterministic. Instead, the MNL model assumes that the parameters
are deterministic and ignores the possibility that the choice parameters
may be random. However, although the RC model allows one to examine whether parameters are random or deterministic, we have not formally developed a {statistical} test to do so. In Appendix \ref{AppendixDeterministicParameters}, we illustrate that 
the estimates from the RC model sometimes suggest that parameters are random when they are deterministic.
We leave analyzing why this occurs and {developing (and analyzing) a formal statistical test} for future research.


\section{Case Study: A Sequential Patent Citation Network}
\label{sec:Citation Network}

We now apply the RC model to a large sequential network that we construct from empirical data. Recall that a network is `sequential' if nodes form edges one after another in a 
sequence.


\subsection{Network Data}
\label{subsec:CiteData}

We apply the RC model to a patent citation network. We construct this network from a subset of a patent data
set that consists of 2,923,922 United States utility patents that were awarded between 1 January 1963
and 31 December 1999. 

We accessed the data from the National Bureau of Economic Research's (NBER) website; see \cite{PatentCitationData} for a discussion of the data. 
This data set has been used extensively to study innovation; see, for example, \cite{aghion2013innovation, bloom2013identifying}. Leskovec \emph{et al.} \cite{leskovec2007graph} used this data set to study how several network properties change with time.

In addition to the citation data, this data set also includes additional information
about each patent. This information includes the award year, the country or countries of its inventor(s), the technology
category, the number of patents that it cites (`citations made'), and the number of patents that cited it (`citations received') \cite{PatentCitationData}. 
In our case study, we consider the `choices' of patents that were awarded in 1975. We focus on 1975 because it is the first year for which have citation data. In Section \ref{subsec:CiteDifferent-Years}, we examine patents that were awarded in other years.

We study the growth of the patent citation network as new patents are awarded in 1975. Each node in the network is a patent; we use the term `new patent' for any patent that was awarded in 1975. This network also includes all patents that were granted before 1975. In particular, although we include all citations of pre-1975 patents in the network, we do not study the `choices' of such patents. Each new patent `chooses' to cite previous patents; there is a directed,
unweighted edge from node $u$ to node $v$ if the new patent $u$ `chooses'
to cite the existing patent $v$.  The new nodes arrive sequentially over time and form edges to existing nodes. Each node can form edges to any node that corresponds to a patent from 1974 or earlier. (We do not allow 1975 patents to choose other nodes from 1975.) (Fewer than $9$\% of the 1975 patents cited another patent from 1975.)

The mean number of citations by each new patent
is $4.875$, and the median number of citations is $4$. 
About 21.3\% of the new patents cite only $1$ patent, so most new patents cite multiple existing patents.
Consequently, the single-choice assumption of the MNL is violated, so it seems sensible to examine this network using the RC model.

In the patent citation network,  there are 69,323 new patents in 1975 that cited at least $1$
patent from 1974 or earlier. There are 636,444 patents that were awarded prior to 1975. In the RC model, it is computationally expensive to estimate parameters, and the computational cost increases with the number of nodes. The number of new patents in 1975 is too large for us to use the entire network for estimation, so we apply a variety of sampling procedures to make estimation computationally tractable.

Of the 69,323 new patents in 1975, we sample 10,000 new patents uniformly at random to form a sample network. As we described in Section
\ref{subsec:Repeated-Choice-Model}, each node (that is, each new patent) has an associated choice sequence. Recall that each choice in the choice sequence consists of a `chosen' alternative and an alternative set.

We assume that the alternative set consists of all previously awarded patents.\footnote{The authors of a new patent presumably only consider a subset of previously existing patents. See \cite{konig2016formation} for a model of network formation with partial observability.} Each new patent can cite any of the previously
awarded patents, so the size of its alternative set is equal to the total number of patents that were awarded prior to 1975. (Recall that we are ignoring citations of patents that were awarded in 1975.) Because 636,444 patents were awarded prior to 1975, the alternative set is very large; we need to reduce its size for computational tractability.

Hall \emph{et al.} \cite{PatentCitationData} assigned each patent to a technology category:
chemical (excluding drugs); computers and communications (C \& C);
drugs and medical (D \& M); electrical and electronics (E \& E); mechanical;
and others. Hall \emph{et al.} \cite{PatentCitationData} further divided each category into sub-categories. For example, the category `Drugs and Medical' was further subdivided into `Drugs', `Surgery and Medical Instruments', `Biotechnology', and `Miscellaneous-Drug\&Med'. For additional details, see Hall \emph{et al.} \cite{PatentCitationData}. On average, approximately 80\% of all of the patents that are cited by a new patent in 1975 are in the same technology category as that
new patent. Therefore, to narrow the alternative set, we consider
only patents in the same technology category as potential 
alternatives. Although we make our assumption for convenience, {the local-search model in \cite{jackson2010social}} provides some theoretical justification for this sampling choice. The basic idea behind this local-search model is that searching for potential nodes with which to link (i.e., searching for patents to cite in the present case study) is `costly'. Consequently, nodes may only consider other nodes that are inexpensive to `find'. In our example, we are assuming that patents in a different technology category from the `new' patent are too expensive to `find'. Jaffe and Trajtenberg \cite{jaffe1999international} concluded that patents are about $100$ times more likely to cite patents in the same technology category than patents in a different category, so this simplification seems reasonable. In Section \ref{subsec:CiteDifferent-Sampling-Methods}, we demonstrate that our results are robust to violations of this assumption.

In each alternative set, we uniformly-at-random sample 6 patents from all of the patents that were not chosen.\footnote{This procedure is often called `negative sampling' \cite{Train_Discrete_Book}. See \cite{overgoor2020} for a recent discussion of negative sampling for relational social data.} This reduces the size
of the alternative set to 7, as there are 6 patents that are not cited and $1$ patent that is cited. We do this sampling because of computational constraints, rather than because of any belief that this sampling satisfies any realistic assumption. This sampling reduces the size of the data very substantially. As was shown in \cite{McFadden_Uniform_Conditioning_Property} and employed in Overgoor \emph{et al.} \cite{DiscreteChoice}, as long as a sampling method
satisfies the property of uniform conditioning\footnote{{Uniform conditioning is satisfied when the choice set consists of the chosen alternative and one or more other randomly selected alternatives \cite{McFadden_Uniform_Conditioning_Property}.}} (which is true of sampling uniformly at random), one still obtains estimates that are statistically consistent. In Section \ref{subsec:CiteRobustness}, we analyze the robustness of our results to different sampling methods.


\subsection{Model Specification}
\label{subsec:CiteModel-Specification}

We specify a model in which the citation probability depends on three observable
factors: citations received (i.e., in-degree), sub-category, and time difference (as measured when the new patent was awarded).

An important assumption of the RC model is that the edge-formation choices depend on observable factors that are deterministic when a choice is made (see Section \ref{subsec:Repeated-Choice-Model}). In the present case study, we are able to study the effect of in-degree on future edge formation because our patent citation network grows sequentially. We assume that nodes make their citation decisions one at a time. The in-degree of an alternative for a given node may depend on the choices that were made by nodes that arrived previously; however, because those choices have already been made, each node's in-degree is deterministic. This allows us to use the RC model to study the effect that the in-degree (which is specified at one time and then remains constant) has on edge formation. This implies that edge formation in the patent citation network is not independent. The sequential nature of our patent citation network allows us to study edge-formation choices without assuming edge independence. Our assumption of sequential node arrival in the present case study is not required to use the RC model.

We normalize the number of citations that are received by a new patent by dividing this number by the number
of patents that are cited by the new patent. For computational convenience, we assume that the choice parameter
for citations received follows a log-normal distribution. Our assumption
is guided by results from previous studies, which suggested that previous
citations have a non-negative effect on the probability of new citations
\cite{jeong2003measuring,newman2009}. In other words, having more citations never hurts a patent's chances of being cited again (although
it may not help). There are several distributions that have a non-negative support, and a log-normal one has convenient properties.\footnote{Specifically, it is easy to compute the unconditional probability in Equation {\eqref{eq:expected_prob}}.} The notion that a node's degree affects future edge formation is a major focus of most PA models \cite{newman2010networks}, and it seems relevant for our case study. 

The binary variable `Sub-Category' indicates whether or not a new patent and a cited patent belong to the same technology sub-category. Recall that we use the category level of classification to construct the alternative set \eqref{subsec:CiteData}. We use the sub-category level of classification as a variable that affects the choice probability. The idea that patents prefer to cite similar patents is conceptually similar to homophily. We assume that the choice parameter is log-normal because it is computationally convenient.

The time difference is equal to the award year of the new patent minus the award year of the cited patent. We assume that the choice parameter for time difference is normally distributed. The normal distribution is computationally convenient, and it allows the age of a patent to affect citation probability either positively or negatively. 

Combining the above considerations yields the utility function
\begin{equation}\label{eg:patent_utility}
	V_{njt}=\beta_{1n}\,\textrm{CReceived\ensuremath{_{njt}}} + \beta_{2n}\,\textrm{SubCat}{}_{njt} + \beta_{3n}\,\textrm{TimeDiff}_{njt}+\epsilon_{njt}\,,
\end{equation}
where `CReceived' is the number of citations received, `SubCat' is a Boolean
variable that is true if both patents are in the same technology sub-category,
and `TimeDiff' is the difference in the award years of the new patent
and the cited patent. The subscript $njt$ denotes the $j^{\textrm{th}}$
alternative of the $t^{\textrm{th}}$ choice in node $n$'s choice
sequence. From our assumptions, the choice parameters are $\beta_{1n}\sim\textrm{LogNormal}(m_{1},s_{1}^{2})$, $\beta_{2n}\sim\textrm{LogNormal}(m_{2},s_{2}^{2})$, and $\beta_{3n}\sim\mathcal{N}(m_{3},s_{3}^{2})$. We draw $\epsilon_{njt}$ from an IID Gumbel distribution.

We allow correlations between different choice parameters and report the correlation parameters. We denote the correlation between choice parameters $\beta_{in}$ and $\beta_{jn}$ by $\textrm{cor}_{ij}$.

We estimate the parameters $m_{1}$, $m_{2}$, $m_{3}$, $s_{1}$, $s_{2}$,
$s_{3}$, $\textrm{cor}_{12}$, $\textrm{cor}_{13}$, and $\textrm{cor}_{23}$ using the {\sc mlogit} package in {\tt R}. We described our estimation procedure, which took approximately 2 hours, in Section \ref{subsec:SynEstimatiom}.


\subsection{Results}
\label{subsec:CiteResults}

\begin{table}[H]
\caption{\textbf{{Results from the RC model for the 1975 patent
citation network.}}
{The first column specifies the parameters. The second column specifies the associated characteristics. For example, the parameter $m_{1}$ is the mean of the distribution of $\beta_{1n}$. As defined in Equation {\eqref{eg:patent_utility}}, the amount that $\textrm{CReceived}_{njt}$ affects the choice utility depends on $\beta_{1n}$. 
The third column gives the estimated values of the parameters, the
fourth column gives the standard errors of these estimates, and the
last column gives the p-values. For statistical-significance levels that are larger than the p-value,
the coefficient is significantly different from $0$. `CReceived' is the number of citations received, `SubCat' is a Boolean
variable that is true if both patents are in the same technology sub-category,
and `TimeDiff' is {the} award year of the new patent minus the award year of the cited patent.}
}
\label{tab:CiteResults-from-MX}

\vspace{1cm}

\begin{centering}
\begin{tabular}{p{1.5cm} p{1.5cm}  p{1.5cm}  p{2.5cm}  p{2.5cm} }
\hline 
Parameter & Characteristic &Coefficient & {Standard Error} & p-value\tabularnewline
\hline 
\hline 
$m_{1}$ & $\textrm{CReceived}$ & $-1.4749$ & 0.0269 & $<0.0001^{***}$\tabularnewline
\hline 
$s_{1}$ & $\textrm{CReceived}$ & 1.5329 & 0.0096 & $<0.0001^{***}$\tabularnewline
\hline 
$m_{2}$ & $\textrm{SubCat}$ & 1.0956 & 0.0121 & $<0.0001^{***}$\tabularnewline
\hline 
$s_{2}$ & $\textrm{SubCat}$ & 0.1017 & 0.0063 & $<0.0001^{***}$\tabularnewline
\hline 
$m_{3}$ & $\textrm{TimeDiff}$ & $-0.0090$ & 0.0031 & 0.0042$^{**}$\tabularnewline
\hline 
$s_{3}$ & $\textrm{TimeDiff}$ & 1.3045 & 0.0245 & $<0.0001^{***}$\tabularnewline
\hline 
$\mathrm{cor}_{12}$ & - & $-0.1730$ & 0.1064 & 0.1042\tabularnewline
\hline 
$\mathrm{cor}_{23}$& -  & 0.2058 & 0.2426 & 0.3962\tabularnewline
\hline 
$\mathrm{cor}_{13}$& -  & $-0.0275$ & 0.02756 & 0.3188\tabularnewline
\hline 
\end{tabular}
\par\end{centering}
\vspace{0.5cm}

\textquoteleft {*}{*}{*}\textquoteright, \textquoteleft {*}{*}\textquoteright,
\textquoteleft {*}\textquoteright, and \textquoteleft {$\bullet$}'
represent statistical-significance levels of {less than 0.01\%}, precisely 0.1\%, 1\%, and 5\%, respectively.
The absence of asterisks or \textquoteleft {$\bullet$}' indicates
that the parameters are not significantly different from $0$.
\end{table}

We summarize our results in Table \ref{tab:CiteResults-from-MX}.
These estimates indicate that $\beta_{1n}\sim\textrm{LogNormal}(-1.4749,1.5329^{2})$, $\beta_{2n}\sim\textrm{LogNormal}(1.0956,0.1017^{2})$, and $\beta_{3n}\sim\mathcal{N}(-0.0090,1.3405^{2})$. All of the estimates for the distribution parameters are statistically
significant. Moreover, there is a strong correlation between
attachment probability and the factors that we are studying. 
The correlation parameters are not statistically significant
and they have small magnitudes, so any correlations between the choice parameters appear to be weak ones. See Section \ref{subsec:Cite-Correlation-Analysis} for a detailed discussion of the correlation parameters. The standard
deviations of all of the choice parameters are statistically different
from $0$, so these parameters are random with a large probability.

Our estimates give distributions of the choice parameters, but these distributions
are difficult to interpret directly. We interpret the results using
different methods in Sections \ref{subsec:CiteInferenceMedian}, \ref{subsec:CiteConfidence-Intervals},
and \ref{subsec:Cite-Correlation-Analysis}. 


\subsubsection{Inference: An Example using Median Values}
\label{subsec:CiteInferenceMedian}

We compute median values from the distributions of $\beta_{in}$ to
obtain point estimates of the choice parameters.

\begin{table}[H]
\caption{\textbf{{Median values of the choice parameters for the 1975 patent citation network.}} 
{The first column specifies the parameters. The second column specifies the associated characteristics. For example, the parameter $m_{1}$ is the mean of the distribution of $\beta_{1n}$. As defined in Equation {\eqref{eg:patent_utility}}, the amount that $\textrm{CReceived}_{njt}$ affects the choice utility depends on $\beta_{1n}$.} The third column
gives the median values of the characteristics.
}
\label{tab:CiteMedian-Values}

\vspace{1cm}

\centering{}%
\begin{tabular}{p{1.5cm} p{1.5cm} p{1.5cm} }
\hline 
Parameter & Characteristic & Median\tabularnewline
\hline 
\hline 
$\beta_{1n}$ & $\textrm{CReceived}$ & 0.2287\tabularnewline
\hline 
$\beta_{2n}$ & $\textrm{SubCat}$ & 2.9909\tabularnewline
\hline 
$\beta_{3n}$ & $\textrm{TimeDiff}$ & $-0.009$\tabularnewline
\hline 
\end{tabular}
\end{table}

We interpret the median values of the choice parameters in the same
way that we interpreted the results of the MNL model. For instance, dividing $\beta_{2n}$
by $\beta_{1n}$ yields a `substitution rate' between citations
received and patent sub-category. With our model specification, this represents the number of additional
citations that a patent in a different sub-category from the new
patent needs to have for it to have the same choice probability as a patent from the same category as the new patent. Specifically, $\beta_{2n}/\beta_{1n} \approx 13$ additional
citations received can compensate for not being in the same sub-category
as the new patent. Similarly, receiving $\beta_{3n}/\beta_{1n} \approx 0.03$ additional citations has the
same effect as a one-year decrease in the time difference between a new patent and a cited patent. This suggests
that age has a weak effect on choice probability, whereas citations
received and sub-category have strong effects.


\subsubsection{Inference: Intervals}
\label{subsec:CiteConfidence-Intervals}

The RC model has the potential to be informative about the distributions of choice parameters. However, a key challenge is that a distribution can have a large support. (For example, the support of $\beta_{3n}$ is the set $\mathbb{R}$ of real numbers.) Many of the parameter values in the support occur with a small probability, so they are likely to be of limited interest. One can use the distribution of choice parameters to compute an interval of parameters such that the choice parameter lies in this interval with a large
probability. For instance, one can compute an interval such that the probability that the choice parameter is in this interval is $0.9$.

As an illustration, we construct such an interval for the parameter for time difference. By assumption, the parameter for time difference
is distributed normally, so $\beta_{3n}\sim\mathcal{N}(-0.0090,1.3405^{2})$.
From a direct computation, we obtain $P(-2.278<\beta_{3n}<2.260) \approx 0.9095$.
Therefore, the time-difference parameter value ranges from $-2.2782$
to $2.260$ with a probability of about $0.9$. One can also construct intervals of different precisions.


\subsubsection{Inference: Correlation Analysis}
\label{subsec:Cite-Correlation-Analysis}

We also use the RC model to study correlations between the choice parameters. Such calculations are helpful for examining whether or not a strong affinity for one choice parameter implies a strong affinity for another choice parameter. For instance, do individuals who cite more-popular patents also tend to cite older patents? Under the assumptions of our framework, the answer to this question is `yes' if the correlation between $\beta_{1n}$ and $\beta_{3n}$ is negative, statistically significant, and has value that is practically significant. From Table \ref{tab:CiteResults-from-MX}, we see that the value is negative but is not statistically significant. Therefore, we conclude that the correlation between citations received and time difference does not differ significantly from $0$. To evaluate this hypothesis in a more rigorous way, we conduct a Wald test \cite{fahrmeir2013regression} and a likelihood-ratio test to examine whether or not there is a correlation Both tests imply that the correlation between the variables is statistically significant; assuming that the correlation is $0$ results in worse fit to the data. However, the correlation is small, as indicated by the likelihood-ratio test. Therefore, although we find a correlation between the choice parameters, it is not of practical significance.


\subsubsection{Discussion}

Our results suggest that attachment probability in the patent citation network has a strong statistical association with the factors that we study. The positive association of the variable SubCat with the attachment probability suggests that there is homophily in the network. The positive association of CReceived with the attachment probability has been documented by several papers (see, e.g., \cite{jeong2003measuring,DiscreteChoice}). 

Using the RC model has allowed us to simultaneously examine homophily and PA by in-degree.  
As we discussed in Section \ref{Advantages_DC}, this is a beneficial property of DC models. We are able to statistically study homophily and PA by choosing appropriate parameters in the RC model without having to make large changes in the RC model. This illustrates the flexibility of the RC model that we discussed in Section \ref{Advantages_DC}. Our results for the patent citation network suggest that PA occurs even if we control for homophily (and vice versa).


\subsection{Robustness of Our Results: Different Award Years and Different {{Data-Construction} Procedures}}
\label{subsec:CiteRobustness}

We now examine patent citation networks from different years and using different sampling methods. Specifically, we study the 
citation choices of new patents from different specified years while fixing the patent citations from earlier years.

In Section \ref{subsec:CiteData}, we described how we sampled data to reduce computational
complexity. For example, we considered new patents only from 1975.
In Section \ref{subsec:CiteDifferent-Years}, we examine patent citation networks for different years and compare the inferred parameter values across years.

In our previous calculations, after restricting the year of awarded patents, we sampled our data further to reduce its size.
First, we sampled 10,000 new patents uniformly at random
from all new patents in 1975. Second, we sampled 6 
alternatives (in addition to the chosen alternative) uniformly at random
to construct the alternative set for each choice in each choice sequence.
Third, we only considered patents in the same technology category
as valid alternatives. 

In the last paragraph of Section \ref{subsec:CiteData},
we argued that applying the RC model to this sampled subset of the
data provides statistically consistent parameter estimates. (See the discussions in
\cite{Train_Discrete_Book,McFadden_Uniform_Conditioning_Property}.)
However, there may be qualitatively important differences
in the parameter estimates when considering different years or different sampling
choices. In Section \ref{subsec:CiteDifferent-Sampling-Methods},
we apply the RC to different sampling choices of the data and compare
the resulting parameter estimates.


\subsubsection{Different Years}
\label{subsec:CiteDifferent-Years}

In Table \ref{tab:CiteDiffYears}, we show our parameter estimates
for patent citation networks from 1975, 1980, 1985, 1990, and 1995. We further sample the data in 
same way that we described in Section \ref{subsec:CiteData}, and our model
specification is the same as in Section \ref{subsec:CiteModel-Specification}.

\begin{table}[H]
\caption{\textbf{{Estimated parameters for the patent citation networks from 1975, 1980, 1985, 1990, and 1995.}} 
{The first column specifies the parameters. The second column specifies the associated characteristics. For example, the parameter $m_{1}$ is the mean of the distribution of $\beta_{1n}$. As defined in Equation \eqref{eg:patent_utility}, the amount that $\textrm{CReceived}_{njt}$ affects the choice utility depends on $\beta_{1n}$. The third, fourth, fifth, sixth, and seventh columns give the estimated coefficients for the years 1975, 1980, 1985, 1990, and 1995, respectively.} }
\label{tab:CiteDiffYears}

\vspace{1cm}

\begin{centering}
\begin{tabular}{ l l l l l l l }
\hline 
Parameter & Characteristic & 1975 & 1980 & 1985 & 1990 & 1995\tabularnewline
\hline 
$m_{1}$ & $\textrm{CReceived}$ & $-1.474^{***}$ & $-1.182^{***}$ & $-1.139^{***}$ & $-1.160^{***}$ & $-1.250^{***}$\tabularnewline
\hline 
$s_{1}$ & $\textrm{CReceived}$ & $1.533^{***}$ & $1.521^{***}$ & $1.530^{***}$ & $1.546^{***}$ & $1.436^{***}$\tabularnewline
\hline 
$m_{2}$ & $\textrm{SubCat}$ & $1.096^{***}$ & $1.064^{***}$ & $0.887^{***}$ & $0.817^{***}$ & $0.734^{***}$\tabularnewline
\hline 
$s_{2}$ & $\textrm{SubCat}$ & $0.102^{***}$ & $0.108^{***}$ & $0.084^{***}$ & $0.073^{***}$ & $0.068^{***}$\tabularnewline
\hline 
$m_{3}$ & $\textrm{TimeDiff}$ & $-0.001^{***}$ & $-0.034^{***}$ & $-0.050^{***}$ & $-0.062^{***}$ & $-0.078^{***}$\tabularnewline
\hline 
$s_{3}$ & $\textrm{TimeDiff}$ & $1.305^{***}$ & $1.225^{***}$ & $1.231^{***}$ & $1.227^{***}$ & $1.339^{***}$\tabularnewline
\hline 
$\mathrm{cor}_{12}$& - & $-0.173$ & $-0.097$ & $-0.028$ & $-0.044$ & $-0.008$\tabularnewline
\hline 
$\mathrm{cor}_{13}$& - & $-0.027$ & $0.088^{***}$ & $0.065^{***}$ & $0.013$ & $-0.117^{***}$\tabularnewline
\hline 
$\mathrm{cor}_{23}$& - & $0.206$ & $-0.046$ & $0.050$ & $-0.093$ & $-0.168^{*}$\tabularnewline
\hline 
\end{tabular}
\par\end{centering}
\vspace{0.5cm}

\textquoteleft {*}{*}{*}\textquoteright, \textquoteleft {*}{*}\textquoteright,
\textquoteleft {*}\textquoteright, and \textquoteleft {$\bullet$}\textquoteright{}
represent statistical-significance levels of {less than 0.01\%}, precisely 0.1\%, 1\%, and 5\%, respectively.
The absence of asterisks or \textquoteleft {$\bullet$}\textquoteright{} indicates
that the parameters are not significantly different from $0$.
\end{table}

For each of the patent years that we consider, the distribution parameters are similar in magnitude and have the same statistical-significance level. The distribution of $\beta_{1n}$ is qualitatively similar for all of the examined years. For the examined years,
we observe a decreasing trend in the parameters $m_{2}$
and $m_{3}$. The trend in $m_2$ suggests that the importance of sub-category
has decreased with time. We also observe a decrease in $s_{2}$ with time. 
These observed trends are suggestive of general ones, and examining all years in the patent citation network using temporal network analysis \cite{holme2019} is an important direction for future research.

The correlation ($\textrm{cor}_{12}$) between $\beta_{1n}$ and $\beta_{2n}$
is statistically insignificant for each of the examined patent years. Similarly,
$\textrm{cor}_{23}$ is statistically insignificant for all years except 1995.
The associated correlation parameters are small in magnitude for all of these years.


\subsubsection{Different {Data-Construction Procedures}}
\label{subsec:CiteDifferent-Sampling-Methods}

We now examine the distribution parameters for different numbers
of new patents that we sample from the data, different numbers of alternatives
in the alternative set of each choice, and without the restriction that alternatives must be in the same technology category as
the new patent. We report our results in Table \ref{tab:CiteDiffSampling}.

\begin{table}[H]
\caption{\textbf{{Estimated parameters for the 1975 patent citation network
 for different {data-construction procedures}.}} 
{The first column specifies the parameters. The second column specifies the associated characteristics. For example, the parameter $m_{1}$ is the mean of the distribution of $\beta_{1n}$. As defined in Equation \eqref{eg:patent_utility}, the amount that $\textrm{CReceived}_{njt}$ affects the choice utility depends on $\beta_{1n}$.
In the {procedure} (1) (i.e., in the third column), we give our baseline sampling method with 10,000 new patents
and 6 alternatives. In {procedure} (2), {we sample 10,000 new patents and each new patent has 3 alternatives.}
In {procedures} (3) and (4),
we sample 5,000 and 20,000 new patents, respectively; the number of alternatives is 6.
In {procedure} (5), we use the same sampling method as in {procedure} (1), but we no longer require the alternatives to be patents from the same
technology category as the new patent.}
}
\label{tab:CiteDiffSampling}

\vspace{1cm}

\noindent \begin{centering}
\begin{tabular}{l l l l l l l}
\hline 
 Parameter & Characteristic & (1) & (2) & (3) & (4) & (5)\tabularnewline
\hline 
$m_{1}$ & $\textrm{CReceived}$ & $-1.474^{***}$ & $-1.299^{***}$ & $-1.360^{***}$ & $-1.469^{***}$ & $-1.308^{***}$\tabularnewline
\hline 
$s_{1}$ & $\textrm{CReceived}$ & $1.533^{***}$ & $1.524^{***}$ & $1.496^{***}$ & $1.510^{***}$ & $0.965^{***}$\tabularnewline
\hline 
$m_{2}$ & $\textrm{SubCat}$ & $1.096^{***}$ & $1.274^{***}$ & $1.112^{***}$ & $1.078^{***}$ & $1.963^{***}$\tabularnewline
\hline 
$s_{2}$ & $\textrm{SubCat}$ & $0.102^{***}$ & $0.111^{***}$ & $0.093^{***}$ & $0.101^{***}$ & $0.093^{***}$\tabularnewline
\hline 
$m_{3}$ & $\textrm{TimeDiff}$ & $-0.009^{**}$ & $-0.003$ & $-0.007$ & $-0.005^{*}$ & $-0.005$\tabularnewline
\hline 
$s_{3}$ & $\textrm{TimeDiff}$ & $1.305^{***}$ & $1.281^{***}$ & $1.287^{***}$ & $1.298^{***}$ & $1.032^{***}$\tabularnewline
\hline 
$\mathrm{cor}_{12}$  & - & $-0.173$ & $-0.129$ & $-0.216$ & $-0.160^{*}$ & $-0.127$\tabularnewline
\hline 
$\mathrm{cor}_{13}$ & - & $-0.027$ & $-0.270^{***}$ & $-0.033$ & $-0.339$ & $0.006$\tabularnewline
\hline 
$\mathrm{cor}_{23}$ & - & $0.206$ & $0.145$ & $0.178$ & $0.073$ & $-0.053$\tabularnewline
\hline 
\end{tabular}
\par\end{centering}
\vspace{0.5cm}

\textquoteleft {*}{*}{*}\textquoteright, \textquoteleft {*}{*}\textquoteright,
\textquoteleft {*}\textquoteright, and \textquoteleft {$\bullet$}\textquoteright{}
represent statistical-significance levels of {less than 0.01\%}, precisely 0.1\%, 1\%, and 5\%, respectively.
The absence of asterisks or \textquoteleft {$\bullet$}\textquoteright{} indicates
that the parameters are not significantly different from $0$.
\end{table}

The citations-received, sub-category parameters (i.e., $m_{1}$, $s_{1}$, $m_{2}$, and $s_{2}$),
and the standard deviation of the time difference (i.e., $s_{3}$) are
all statistically significant for all of the {procedures}. In {procedures} (1)--(4) of
Table \ref{tab:CiteDiffSampling}, we observe that the values of these parameters are mostly very similar to each other, although $\textrm{cor}_{13}$ has a larger magnitude in {procedure} (4) than it does {in the other procedures}.
Broadly speaking, the parameter values appear to be robust with respect to the choices in these four {procedures}.

In {procedure} (5) of Table \ref{tab:CiteDiffSampling}, we observe that $m_{1}$, $m_{3}$, $s_{2}$, and $s_{3}$ have similar
magnitudes as {in} the other {procedures}. We also observe that the value of $s_{1}$ {in procedure} (5) has a smaller magnitude than it does {in} the other {procedures}. However, $m_{2}$ has a larger value than {in} the other {procedures}. Because we no longer require the new patent and its alternatives to be in the same technology category, one possible explanation for this difference is that we are now incorporating a notion of {technology-category} relatedness between patents with the sub-category variable. By definition, if two patents are in the same technology sub-category, they must be in the same technology category. However, the converse is not true.
Consequently, the presence of two patents in the same technology sub-category is a stronger measure of relatedness
when we do not control for the technology category. This increases the
relative importance of the sub-category variable and helps explain {why the distribution of $\beta_{2n}$ has a larger mean in {procedure (5) than in the other procedures}.}

The correlation parameters in Table \ref{tab:CiteDiffSampling} are not statistically significant, and they usually have
a small magnitude. (For example, 13 of the 15 parameters have an absolute value that is less than $0.22$.) 
This supports our prior conclusion that the correlation between choice parameters is not of practical significance. The mean parameter for time difference
($m_{3}$) is not statistically significantly different from $0$ in three different {procedures} (see the results for {procedures} (2), (3), and (5)), and it is close
to $0$ in magnitude for all procedures.

Although we {are} not exhaustive in our different {data-construction procedures}, the {examined procedures suggest} that our qualitative interpretation of the parameter
estimates appear to be mostly robust with respect both to differences in the number of new patents that we sample and to our different procedures.


\subsection{Goodness-of-Fit Tests}
\label{goodness}

To evaluate whether or not the RC model accurately captures the structure of a patent citation network, we perform goodness-of-fit tests~\cite{hunter2008goodness, cranmer2020inferential, brandenberger2019predicting}. In a goodness-of-fit test, one generates networks using a fitted model and compares {the structural statistics of an} observed network to those of the synthetic
networks. If the statistics from the synthetic networks align with those of the observed network, then one concludes that the model has successfully fit the data. In the context of our setting, this suggests that a model has generated networks with appropriate statistical properties and that it captures relevant features of the network-formation process (under the assumptions of the goodness-of-fit tests).

Recall that we fit our model using new patents from 1975. To provide an out-of-sample assessment, we examine the citations of the new patents in the first three months of 1976. We refer to this network as the ``early-1976 network". Using our fitted RC model, we generate synthetic edges (i.e., citations of existing patents) for the patents in the early-1976 network. To reduce uncertainty from stochasticity, we create an ensemble of 100 of these synthetic networks. For each network in the ensemble, we compute the distributions of the out-degree, the in-degree, citation homophily by patent sub-category, the number of weakly connected components, and the geodesic distance between nodes. We then aggregate these statistics across the 100 networks in our ensemble and compare these aggregate statistics to the corresponding statistics in the observed (i.e., early-1976) network. 

In Figure \ref{fig:gof}, we present the results of our goodness-of-fit tests. In Figure \ref{fig:gof}A, we show that our synthetic networks closely replicate the out-degree distribution of the early-1976 network. Although the out-degree distribution is a common statistic in goodness-of-fit tests \cite{hunter2008goodness, cranmer2020inferential, brandenberger2019predicting}, it not particularly informative when evaluating the RC model. The RC model assumes that we know the timing of each choice of each node, so the out-degree of the network is an input of the model. It is not something that we seek to infer using the RC model. We compute the out-degree distribution to check that we have correctly implemented our network-generation procedure.
 
We now examine the RC model's goodness of fit by comparing the statistics that it infers to corresponding statistics in the observed early-1976 network. In Figure \ref{fig:gof}B, we compare the in-degree distributions of the observed and synthetic early-1976 networks and observe a reasonable match between them. In Figures \ref{fig:gof}C,D, we show that the RC model is also reasonably successful at capturing the geodesic distances (i.e., the distances of shortest paths) between nodes. Patents must always cite earlier patents, and a geodesic path from one patent to an earlier one yields the shortest lineage between those two patents. In Figure \ref{fig:gof}E, we show that the distribution of the sizes of the weakly connected components is similar in the observed and synthetic networks. Our goodness-of-fit tests also illustrate a flaw in our RC model. In particular, we see in Figure \ref{fig:gof}C that the RC model underestimates sub-category homophily, which we measure as the proportion of patent citations by a new patent that are to patents in the same technology sub-category. Therefore, we see that there are some biases in our application of the RC model to patent citation networks.}

In summary, our goodness-of-fit tests suggest that the RC model does a reasonably good job of capturing some key statistical features (e.g., the in-degree and geodesic-distance distributions) of our patent citation network. Under the assumptions of the employed goodness-of-fit tests \cite{hunter2008goodness, cranmer2020inferential, brandenberger2019predicting}, which are common choices in applications of ERGMs and REMs to network data, the RC model thus appears to produce low-bias results for these features. However, it underestimates homophily by patent sub-category.

\begin{figure}[H]
    \centering
    \includegraphics[scale=0.5]{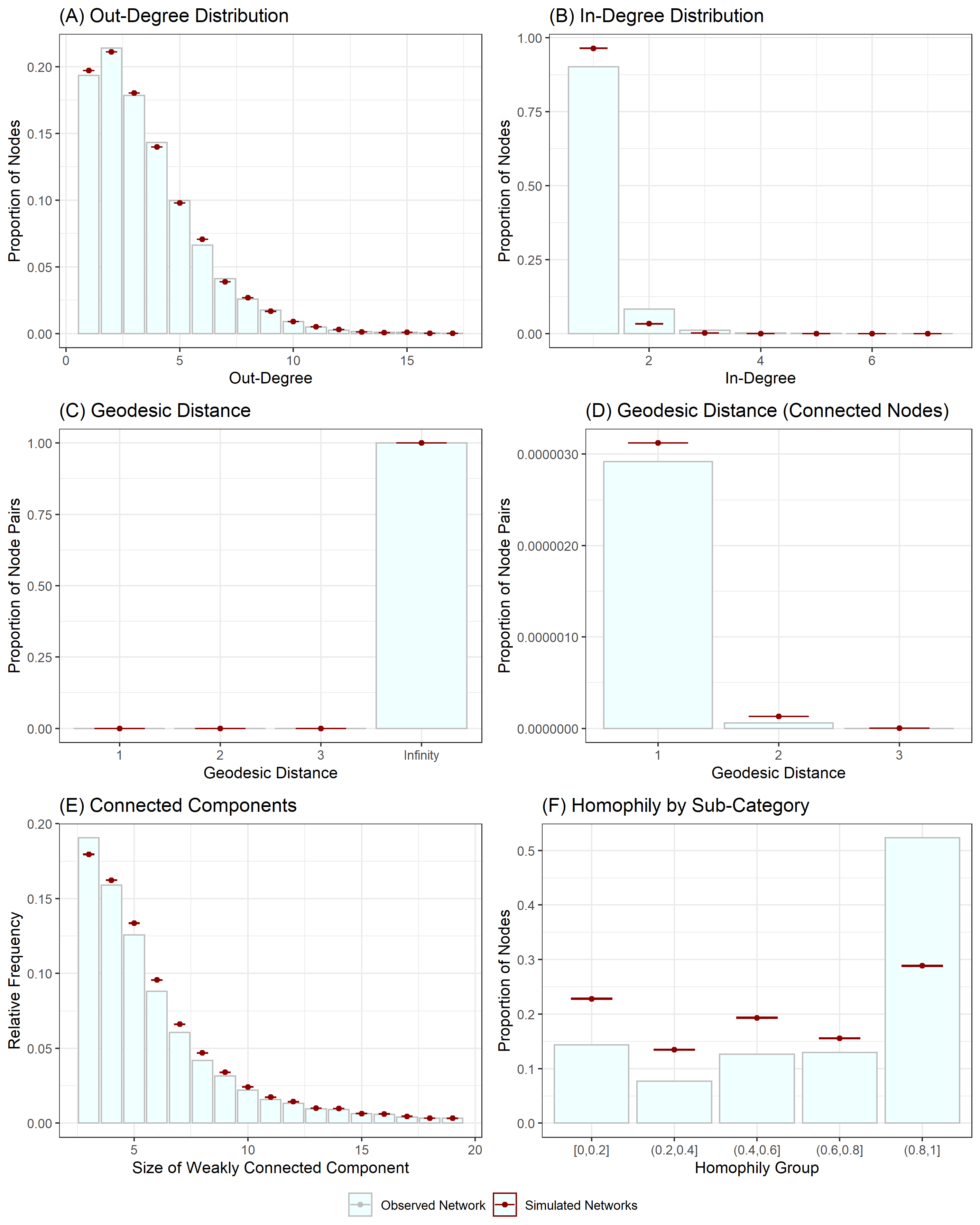}
    \caption{{\bf Goodness-of-Fit tests for several network statistics.}
    {In each panel, the bars show the distribution of a statistic for the observed early-1976 patent citation network and the red points indicate the mean value of that statistic in our
    ensemble of synthetic networks. The red lines show the 95\% confidence intervals for the synthetic networks. Panel (A) shows the distribution the number of citations by new patents (i.e., the out-degree). Panel (B) shows the distribution of the number of citations that are received by each patent in the network (i.e., the in-degree). Panel (C) shows the distribution of the geodesic distances between nodes for all node pairs. The geodesic distance between two nodes is the number of edges in a shortest path between them.
    The geodesic distance from one node to a second node is infinity if there is no path from the first node to the second node.
    Panel (D) shows the distribution of the geodesic distances between nodes for node pairs for which there exists a path. 
    Panel (E) shows the relative frequencies of the size of the weakly connected components of the networks. 
    Panel (F) shows the distribution of the fraction
    of patent citations by a new patent that are to patents in the same technology sub-category (i.e., homophily by sub-category).
   }}
    \label{fig:gof}
\end{figure}


\subsection{Comparison of the RC and MNL Models}
\label{subsec:CiteComparisonMNL}

We compare our results from the RC model to ones from the MNL model.
As we discussed in Section \ref{subsec:CiteData}, our setup violates several assumptions
of the MNL model. However, we ignore those
violations and fit the data to the MNL model to obtain point estimates
for our comparison. Our first observation is that the RC model achieves a much
higher likelihood than the MNL model. Using a likelihood-ratio test, we find a statistically significant difference (with a p-value that is less
than $0.0001$). This highlights that the RC model is a better statistical
fit for our data.

\begin{table}[H]
\caption{\textbf{{Point estimates of the choice parameters from the MNL model for the 1975 patent citation network.}} 
{The first column specifies the parameters. The second column specifies the associated characteristics.
For example, the parameter $m_{1}$ is the mean of the distribution of $\beta_{1n}$. As defined in Equation \eqref{eg:patent_utility}, the amount that $\textrm{CReceived}_{njt}$ affects the choice utility depends on $\beta_{1n}$. The third column gives the estimated values of the parameters, the fourth column gives the standard errors of these estimates, and the last column gives the p-values. For statistical-significance levels that are larger than the p-value,
the coefficient is significantly different from $0$.}
}
\label{tab:CiteMNLEstimates}

\vspace{1cm}

\begin{centering}
\begin{tabular}{p{1.5cm} p{1.5cm} p{1.5cm}  p{2.5cm}  p{2.5cm} }
\hline 
Parameter & Characteristic & Coefficient & {Standard Error} & p-value\tabularnewline
\hline 
\hline 
$\beta_{1}$ & $\textrm{CReceived}$ & 0.2004 & 0.0040 & $<0.0001^{***}$\tabularnewline
\hline 
$\beta_{2}$ & $\textrm{SubCat}$ & 2.1422 & 0.0161 & $<0.0001^{***}$\tabularnewline
\hline 
$\beta_{3}$ & $\textrm{TimeDiff}$ & $-0.0186$ & 0.0022 & $<0.0001^{***}$\tabularnewline
\hline 
\end{tabular}
\par\end{centering}
\vspace{0.5cm}

\textquoteleft {***}\textquoteright, \textquoteleft {**}\textquoteright,
\textquoteleft {*}\textquoteright, and \textquoteleft {$\bullet$}\textquoteright{}
represent statistical-significance level of {less than 0.01\%}, precisely 0.1\%, 1\%, and 5\%, respectively.
\end{table}

The point estimates of the choice parameters in Table
\ref{tab:CiteMNLEstimates} and the median values of the distributions
of the choice parameters in Table \ref{tab:CiteMedian-Values} have similar magnitudes. As we saw in Section \ref{subsec:Random-Choice-Parameters},
this is not always the case. Moreover, there are substantial differences in the interpretation of the results from the RC and MNL models. In particular, in the MNL model, we obtain point estimates, which gives the impression that the choice parameters vary less than is actually the case.

The MNL model ignores the heterogeneity of choice parameters across individuals. 
By treating the choices of an individual across different choices and the choices of two different individuals as the same, the original MNL model
ignores an important aspect of the data-generation process. In reality, these parameters are heterogeneous across nodes and the network growth process is stochastic in two ways. First, the nodes form edges to other nodes through a random process; this feature is captured both by the RC model and by the MNL model.
Second, the importance of the factors that affect attachment probability is different for different nodes.
This second stochastic feature is captured by the RC model, but it is ignored
by the MNL model.

Additionally, as we discussed in Section \ref{subsec:Deterministic-Choice-Parameters},
we can use the RC model to detect whether the choice parameters are
deterministic or random. In our study of a patent citation network (see Table~\ref{tab:CiteDiffSampling}), we found that the standard deviations of all of the choice parameters are different from $0$ with a large probability; this suggests that the choice parameters
are random with a large probability. The MNL provides no information
about this aspect of the data-generation process. Therefore, the RC
model is a more appropriate choice than the MNL model for our case study.


\subsection{Comparison of the RC Model with Other Models for Examining Network Formation}
\label{subsec:cite-comparison-models}

We now briefly compare the RC model with other available models. To help make our comparison concrete, we often make comments in the context of our patent citation network.

The RC model has a lower computational cost than ERGMs, which are popular in the study of network formation.
ERGMs perform estimation based on an MCMC procedure. In an ERGM, the amount of time that a Markov chain needs to converge to a stationary distribution is exponential in the number of nodes of a network unless edge formation is independent (i.e., unless the model is equivalent to an ER model)
\cite{bhamidi2008mixing, chatterjee2013estimating}. It is thus challenging to use an ERGM in a network of the size of our
patent citation network, even after drastically sampling the network as in our investigation, without assuming that edge formation is independent. Some algorithms have been proposed to estimate ERGM parameters in large networks \cite{stivala2020exponential, chakraborty2020patent}, but (as we discuss below) the theoretical properties of those algorithms have not been established. Additionally, sampling to reduce computational burden is more difficult for ERGMs than for {RC models} \cite{handcock2010modeling, pattison2013conditional}. We were able to apply the RC model to study our patent citation network by assuming that the network grows sequentially and using negative sampling. In our case study, we found that previous citations that are received by a patent increase the probability that the patent is cited by a new node, so it is not appropriate to assume edge independence. Recent research on ERGMs has attempted to overcome the computational limitations that we have highlighted. For example, ERGMs were used recently to study subnetworks of a European patent citation network {(this network has about as many nodes as our patent citation network)} by restricting attention to only two patent categories \cite{chakraborty2020patent}. Additionally, Schmid \emph{et al.} \cite{schmid2021generative} used an ERGM to study new citations in a United States Supreme Court citation network in a specified year while fixing the network structure from prior years.

A second key benefit of the RC model is the ability to consider both (1) heterogeneity in choice parameters across nodes and (2) correlations between choice parameters. As we saw in Section \ref{subsec:CiteComparisonMNL}, this is an improvement over the original MNL model, which is in common use. Because many popular models (such as SAOMs and LSMs) use the original MNL model, they are unable to capture this heterogeneity. In theory, many of these models can employ the RC model instead of an MNL model. Exploring these extensions is an avenue for future research. REMs are another popular class of models for the study of sequential networks like our patent citation network \cite{butts20084}, and it is worthwhile to study such networks using an REM that (as in the RC) incorporates heterogeneous and correlated choices parameters of the nodes in a network.

Prior research guarantees that our parameter estimates are both consistent and unbiased~\cite{TrainRevelt97mixedlogit, Train_Discrete_Book}. It is often difficult to establish that a network-formation model is consistent and unbiased. For example, to the best of our knowledge, these properties have not been established for
popular models such as SAOMs and LSMs. Shalizi and Rinaldo~\cite{shalizi2013consistency} showed that projectivity\footnote{{A model is `projective' if one can recover the information from a subnetwork $A$ of a network $G$ from the information of any subnetwork $B$ of $G$ that is a superset of $A$ by applying a `projection mapping'. Intuitively, a projection mapping is a function that drops the `extra data'.}} is a sufficient condition for consistency for any distribution in the exponential family. 
They also showed that any ERGM with a sufficient statistic\footnote{The term `sufficient statistics' refers to a set of network statistics such that no other statistics provide additional information about the parameters of a model.} that is a function of the number of 3-cliques in a network violates this condition. Projectivity is a sufficient condition but not a necessary condition, and there do exist consistency results for non-projective ERGMs~\cite{schweinberger2020exponential}. However, likelihood-based inference for non-projective ERGMs is computationally expensive. 
It is difficult to be precise about the maximum network size that is practical for likelihood-based inference.
Typically, if a network has more than a few thousand nodes, it is difficult to do such inference without making simplifying assumptions. ERGMs have been used  to study networks with {thousands of nodes (see, e.g., \cite{ babkin2020large, schmid2021generative})}, and recent advances promise the ability to use ERGMs to study even larger networks
\cite{krivitsky2022ergm, schmid2017exponential}. Currently, to apply non-projective ERGMs to larger networks, it is necessary either to use inference methods with known problems (e.g., maximum pseudo-likelihood estimation (MPLE) methods). ERGMs seek to capture general features of network formation , and (despite their computational burden) they are useful for many applications, including for complex processes. In the present work, we focus on a relatively simple setting, and the RC is very effective for it.


\section{Limitations and Applicability}
\label{sec:Limitations-and-Applicability}

In this section, we highlight several limitations of the RC model and our analysis. 

First, recall that we assumed that the choice parameters follow a particular distribution type. We cannot test this
assumption because the true distribution is unknown in most situations. Consequently, we risk misspecifying these distributions.
However, as we pointed out in Section \ref{subsec:SDatanadModel},
this limitation is not unique to our model. It is common to make distributional assumptions in DC models \cite{kleinbaum2002logistic,Train_Discrete_Book}.
For example, the MNL model assumes implicitly that its parameters
are deterministic \cite{Train_Discrete_Book}. We also needed to make distributional assumptions to use the RC model.

Another major limitation of the RC model is the computational complexity of using it for estimating distribution parameters. In our computations, the RC model took approximately ten times longer than the MNL model to estimate the relevant parameters.
This computational complexity has two important implications. First, we cannot use the RC model to study large networks without drastic sampling (see Section \ref{subsec:CiteData}). Second, we also need to make `convenient' choices of parameter distribution because the computational complexity depends on the distributions of the choice parameters. The number of convenient distributions may also be limited, as many researchers desire to use existing statistical packages; most such packages only allow parameters to have normal, log-normal, uniform, or triangular distributions \cite{Croissant_estimationof,Stata_MX}. 
Given this convenience, we choose to use these distributions for our computations with the RC model.
In theory, the RC model is applicable using other distributions, but it will be helpful for them to be implemented in statistical packages.

In the RC model, we obtain distributions of the choice parameters instead
of point estimates of them. Unfortunately, it is difficult
to directly interpret the qualitative nature of these distributions. 
In Sections \ref{subsec:CiteInferenceMedian} and \ref{subsec:CiteConfidence-Intervals}, we examined different approaches
for interpreting distributions of the choice parameters.

The limitations that we have just discussed and the assumptions that we discussed in Section \ref{subsec:Repeated-Choice-Model} inform us about the realms of applicability of the RC model. We posit that the RC model is particularly appropriate for directed, unweighted networks in which a large number of nodes have an out-degree that is larger than $1$, the preferences are heterogeneous across nodes, and the choice set is heterogeneous across nodes. These features are common to many networks \cite{newman2010networks}. The RC model is also particularly appropriate for networks in which it is possible to construct the choice set of each node.  Networks in which nodes appear in a sequential order (see Section \ref{sec:Citation Network}) are examples of such networks. In the RC model, we assumed that edge formation is affected by deterministic features, and we expect the RC model to be applicable when the features that we seek to study are plausibly deterministic.


\section{Conclusions and Discussion}
\label{sec:Conclusions-and-Discussions}

We employed discrete-choice (DC) models to study network formation. We argued that the repeated-choice (RC) mixed logit (MX) model is more appropriate than the popular multinomial logit (MNL) model for studying network formation, and we illustrated this claim with examinations of synthetic and empirical networks. A key difference between the MNL model and the RC model is that the RC model allows heterogeneous choice parameters.

We also highlighted several characteristics of the RC model that make it well-suited to studying network formation. In particular, the RC model can account for heterogeneity in the `preferences' of nodes that are associated with a network's structure. The existing literature about DC models provides tools for estimation, testing, and parameter selection in the RC model. Although we focused on the RC model, it is important to note that other extensions of the MNL model may also be well-suited to studying network formation. Examining the potential suitability of these extensions is an avenue for future work.

In our study of the RC model, we examined both synthetic and empirical networks. We applied the RC model to synthetic networks in the form of a set of random networks. In this case study, we illustrated that the RC model can produce accurate estimates of either homogeneous or heterogeneous
preferences of the nodes of a network. We also applied the RC model to a large patent citation network, and we compared our results from the RC model with those from the MNL model. We also compared results from a variety of sampling procedures, which we needed to employ because of the significant computational cost of the RC model. In this comparison, we obtained qualitative results that are largely robust with respect to these different sampling procedures. 

The RC model also complements existing models of network formation. For example, in SAOMs, researchers have examined whether or not an edge forms at each event using the MNL model \cite{snijders2010introduction, snijders2017stochastic}. It may be possible to improve the performance of SAOMs by employing the RC model instead of the MNL model in its choice component. Exploring such applications of the RC model is an avenue for further research.

In Section \ref{subsec:Deterministic-Choice-Parameters}, we discussed a strategy to determine whether the choice parameters in the RC model are random or deterministic. We discussed this strategy intuitively but did not develop a formal statistical test. Moreover, our approach may yield incorrect results (see Appendix \ref{AppendixDeterministicParameters}). Investigating these errors and developing a formal statistical test is an important avenue for future research.

An important limitation of the RC model is that it is computationally expensive. In the present paper, we used `convenient' distributions of choice parameters and sampled large networks very drastically to be able to study them. Future research in computation and simulation methods is necessary to be able to employ a larger set of distributions of choice parameters and to study large networks without such drastic sampling. Overgoor \emph{et al.} \cite{overgoor2020} recently proposed a technique to improve the computational efficiency of MX models. Another potential approach is to attempt to use Gibbs sampling \cite{lynch2007introduction} to estimate the posterior distributions of the choice parameters. 

Attachment models are a popular and important type of network model \cite{newman2010networks}, and it is worth further developing approaches like the present one to obtain insights into the properties of networks that are produced by such generative models. Given empirical network data, it will be particularly interesting to try to use such approaches to help infer which mechanisms may have produced it.


\section*{Acknowledgements}
\label{sec:Acknowledgments}

We thank Unchitta Kanjanasaratool for her contributions to the early stages of this research project, which started as a class project by HG and her. We thank Austin Benson, Arun Chandrashekhar, Michelle Feng, Matthew Jackson, Jay Lu, Nina Otter, and the anonymous referees for their many helpful comments. We also thank the participants at the Student Workshop --- Econometrics and Matthew Jackson's discussion group at Stanford University for their valuable feedback.



\begin{thebibliography}{00}

\bibitem{aghion2013innovation}
Aghion, P., Van~Reenen, J. {\&} Zingales, L. (2013)  Innovation and
  institutional ownership. {\em American Economic Review}, \textbf{103}(1),
  277--304.

\bibitem{albert2002statistical}
Albert, R. {\&} Barab{\'a}si, A.-L. (2002)  Statistical mechanics of complex
  networks. {\em Reviews of Modern Physics}, \textbf{74}(1), 47--97.

\bibitem{babkin2020large}
Babkin, S., Stewart, J., Long, X. {\&} Schweinberger, M. (2020)  Large-scale
  estimation of random graph models with local dependence. {\em Computational
  Statistics \& Data Analysis}, \textbf{152}, 107029.

\bibitem{BensonIIAViolation}
Benson, A.~R., Kumar, R. {\&} Tomkins, A. (2016)  On the Relevance of
  Irrelevant Alternatives. In {\em Proceedings of the 25th International
  Conference on World Wide Web}, WWW '16, pages 963--973, Republic and Canton
  of Geneva, Switzerland.

\bibitem{bhamidi2008mixing}
Bhamidi, S., Bresler, G. {\&} Sly, A. (2011)  Mixing time of exponential random
  graphs. {\em Annals of Applied Probability}, \textbf{21}(6), 2146--2170.

\bibitem{Bianconi_2001}
Bianconi, G. {\&} Barab{\'{a}}si, A.-L. (2001)  Competition and multiscaling in
  evolving networks. {\em Europhysics Letters ({EPL})}, \textbf{54}(4),
  436--442.

\bibitem{blochjackson2006definitions}
Bloch, F. {\&} Jackson, M.~O. (2006)  Definitions of equilibrium in network
  formation games. {\em International Journal of Game Theory}, \textbf{34}(3),
  305--318.

\bibitem{bloom2013identifying}
Bloom, N., Schankerman, M. {\&} Van~Reenen, J. (2013)  Identifying technology
  spillovers and product market rivalry. {\em Econometrica}, \textbf{81}(4),
  1347--1393.

\bibitem{blume2011identification}
Blume, L.~E., Brock, W.~A., Durlauf, S.~N. {\&} Ioannides, Y.~M. (2011)
  Identification of social interactions. In {\em Handbook of Social Economics},
  volume~1, pages 853--964. Elsevier, Amsterdam, The Netherlands.

\bibitem{brandenberger2018trading}
Brandenberger, L. (2018)  Trading favors --- {E}xamining the temporal dynamics
  of reciprocity in congressional collaborations using relational event models.
  {\em Social Networks}, \textbf{54}, 238--253.

\bibitem{brandenberger2019predicting}
Brandenberger, L. (2019)  Predicting network events to assess goodness of fit
  of relational event models. {\em Political Analysis}, \textbf{27}(4),
  556--571.

\bibitem{buhlmann2011statistics}
B{\"u}hlmann, P. {\&} Van De~Geer, S. (2011) {\em Statistics for
  High-Dimensional Data: {M}ethods, Theory and Applications}.
Springer-Verlag, Heidelberg, Germany.

\bibitem{butts20084}
Butts, C.~T. (2008)  A relational event framework for social action. {\em
  Sociological Methodology}, \textbf{38}(1), 155--200.

\bibitem{chakraborty2020patent}
Chakraborty, M., Byshkin, M. {\&} Crestani, F. (2020)  Patent citation network
  analysis: {A} perspective from descriptive statistics and ERGMs. {\em PloS
  ONE}, \textbf{15}(12), e0241797.

\bibitem{chandrasekhar2016econometrics}
Chandrasekhar, A. (2016)  Econometrics of network formation. In {\em The Oxford
  Handbook of the Economics of Networks}, pages 303--357. Oxford University
  Press, Oxford, UK.

\bibitem{chatterjee2013estimating}
Chatterjee, S. {\&} Diaconis, P. (2013)  Estimating and understanding
  exponential random graph models. {\em The Annals of Statistics},
  \textbf{41}(5), 2428--2461.

\bibitem{christakis2010empirical}
Christakis, N., Fowler, J., Imbens, G.~W. {\&} Kalyanaraman, K. (2020)  An
  empirical model for strategic network formation. In {\em The Econometric
  Analysis of Network Data}, pages 123--148. Elsevier, Amsterdam, The
  Netherlands.

\bibitem{cranmer2020inferential}
Cranmer, S.~J., Desmarais, B.~A. {\&} Morgan, J.~W. (2020) {\em Inferential
  Network Analysis}.
Cambridge University Press, Cambridge, UK.

\bibitem{Croissant_estimationof}
Croissant, Y. (2012)  Estimation of multinomial logit models in {R}: The {\sc
  mlogit} packages. {\em CRAN}.
{\tt R} package version 0.2-2. Available at {\tt
  http://cran.r-project.org/web/packages/mlogit/vignettes/mlogit.pdf}.

\bibitem{Price76ageneral}
{de Solla Price}, D. (1976)  A general theory of bibliometric and other
  cumulative advantage processes. {\em Journal of the American Society for
  Information Science}, \textbf{27}(5), 292--306.

\bibitem{erdos59a}
Erd\H{o}s, P. {\&} R\'{e}nyi, A. (1960)  On the evolution of random graphs.
  {\em Institute of Mathematics, Hungarian Academy of Sciences}, \textbf{5}(1),
  17--60.

\bibitem{fafchamps2007risk}
Fafchamps, M. {\&} Gubert, F. (2007)  Risk sharing and network formation. {\em
  American Economic Review}, \textbf{97}(2), 75--79.

\bibitem{fahrmeir2013regression}
Fahrmeir, L., Kneib, T., Lang, S. {\&} Marx, B. (2013) {\em Regression: Models,
  Methods and Applications}.
Springer-Verlag, Berlin, Germany.

\bibitem{fellows2018new}
Fellows, I.~E. (2018)  A new generative statistical model for graphs: {T}he
  latent order logistic ({LOLOG}) model. \textbf{arXiv:1804.04583}.

\bibitem{goldenberg2010survey}
Goldenberg, A., Zheng, A.~X., Fienberg, S.~E. {\&} Airoldi, E.~M. (2010)  A
  survey of statistical network models. {\em Foundations and Trends in Machine
  Learning}, \textbf{2}(2), 129--233.

\bibitem{goldsmith2013social}
Goldsmith-Pinkham, P. {\&} Imbens, G.~W. (2013)  Social networks and the
  identification of peer effects. {\em Journal of Business \& Economic
  Statistics}, \textbf{31}(3), 253--264.

\bibitem{graham2015methods}
Graham, B.~S. (2015)  Methods of identification in social networks. {\em Annual
  Review of Economics}, \textbf{7}(1), 465--485.

\bibitem{graham2014a}
Graham, B.~S. (2017)  An econometric model of network formation with degree
  heterogeneity. {\em Econometrica}, \textbf{85}(4), 1033--1063.

\bibitem{PatentCitationData}
Hall, B.~H., Jaffe, A.~B. {\&} Trajtenberg, M. (2001)  The {N.B.E.R.} patent
  citation data file: {L}essons, insights and methodological tools. Working
  Paper 8498, National Bureau of Economic Research.

\bibitem{handcock2010modeling}
Handcock, M.~S. {\&} Gile, K.~J. (2010)  Modeling social networks from sampled
  data. {\em The Annals of Applied Statistics}, \textbf{4}(1), 5.

\bibitem{handcock2004likelihood}
Handcock, M.~S. {\&} Jones, J.~H. (2004)  Likelihood-based inference for
  stochastic models of sexual network formation. {\em Theoretical Population
  Biology}, \textbf{65}(4), 413--422.

\bibitem{hanks_just_wansink_2012}
Hanks, A.~S., Just, D.~R. {\&} Wansink, B. (2012)  Trigger foods: {T}he
  influence of irrelevant alternatives in school lunchrooms. {\em Agricultural
  and Resource Economics Review}, \textbf{41}(1), 114--123.

\bibitem{hoff2002latent}
Hoff, P.~D., Raftery, A.~E. {\&} Handcock, M.~S. (2002)  Latent space
  approaches to social network analysis. {\em Journal of the American
  Statistical Association}, \textbf{97}(460), 1090--1098.

\bibitem{Stata_MX}
Hole, A.~R. (2007)  {Fitting mixed logit models by using maximum simulated
  likelihood}. {\em Stata Journal}, \textbf{7}(3), 388--401.

\bibitem{holme2019}
Holme, P. {\&} Saram\"{a}ki, J., editors (2019) {\em Temporal Network Theory}.
Springer International Publishing, Cham, Switzerland.

\bibitem{HosmerLemesbow}
Hosmer, D.~W. {\&} Lemesbow, S. (1980)  Goodness of fit tests for the multiple
  logistic regression model. {\em Communications in Statistics --- Theory and
  Methods}, \textbf{9}(10), 1043--1069.

\bibitem{hunter2008goodness}
Hunter, D.~R., Goodreau, S.~M. {\&} Handcock, M.~S. (2008)  Goodness of fit of
  social network models. {\em Journal of the American Statistical Association},
  \textbf{103}(481), 248--258.

\bibitem{jackson2005survey}
Jackson, M.~O. (2005)  A survey of network formation models: {S}tability and
  efficiency. {\em Group Formation in Economics: Networks, Clubs, and
  Coalitions}, \textbf{664}, 11--49.

\bibitem{jackson2010social}
Jackson, M.~O. (2010) {\em Social and Economic Networks}.
Princeton University Press, Princeton, New Jersey, USA.

\bibitem{jaffe1999international}
Jaffe, A.~B. {\&} Trajtenberg, M. (1999)  International knowledge flows:
  {E}vidence from patent citations. {\em Economics of Innovation and New
  Technology}, \textbf{8}(1--2), 105--136.

\bibitem{jeong2003measuring}
Jeong, H., N{\'e}da, Z. {\&} Barab{\'a}si, A.-L. (2003)  Measuring preferential
  attachment in evolving networks. {\em Europhysics Letters ({EPL})},
  \textbf{61}(4), 567--572.

\bibitem{kleinbaum2002logistic}
Kleinbaum, D.~G. {\&} Klein, M. (2002) {\em Logistic Regression}.
Springer-Verlag, Heidelberg, Germany.

\bibitem{konig2016formation}
K{\"o}nig, M.~D. (2016)  The formation of networks with local spillovers and
  limited observability. {\em Theoretical Economics}, \textbf{11}(3), 813--863.

\bibitem{krivitsky2022ergm}
Krivitsky, P.~N., Hunter, D.~R., Morris, M. {\&} Klumb, C. (2022)  {ERGM 4:
  C}omputational Improvements. \textbf{arXiv:2203.08198}.

\bibitem{leskovec2007graph}
Leskovec, J., Kleinberg, J. {\&} Faloutsos, C. (2007)  Graph evolution:
  {D}ensification and shrinking diameters. {\em ACM Transactions on Knowledge
  Discovery from Data (TKDD)}, \textbf{1}(1), 2--es.

\bibitem{lynch2007introduction}
Lynch, S.~M. (2007) {\em Introduction to Applied Bayesian Statistics and
  Estimation for Social Scientists}.
Springer-Verlag, Heidelberg, Germany.

\bibitem{McFadden_Uniform_Conditioning_Property}
McFadden, D. (1978)  Modeling the choice of residential location. {\em
  Transportation Research Record}, \textbf{673}, 72--77.

\bibitem{McFaTrai00}
McFadden, D. {\&} Train, K. (2000)  Mixed {MNL} models for discrete response.
  {\em Journal of Applied Econometrics}, \textbf{15}(5), 447--270.

\bibitem{mele2017structural}
Mele, A. (2017)  A structural model of dense network formation. {\em
  Econometrica}, \textbf{85}(3), 825--850.

\bibitem{menard2002applied}
Menard, S. (2002) {\em Applied Logistic Regression Analysis}.
SAGE Publishing, Thousand Oaks, CA, USA.

\bibitem{newman2009}
Newman, M. E.~J. (2009)  The first-mover advantage in scientific publication.
  {\em Europhysics Letters ({EPL})}, \textbf{86}(6), 68001.

\bibitem{newman2010networks}
Newman, M. E.~J. (2018) {\em Networks}.
Oxford University Press, Oxford, UK, 2nd edition.

\bibitem{DiscreteChoice}
Overgoor, J., Benson, A. {\&} Ugander, J. (2019)  Choosing to grow a graph:
  {M}odeling network formation as discrete choice. In {\em The World Wide Web
  Conference}, WWW '19, pages 1409--1420. Association for Computing Machinery.

\bibitem{overgoor2020}
Overgoor, J., Pakapol~Supaniratisai, G. {\&} Ugander, J. (2020)  Scaling choice
  models of relational social data. In {\em Proceedings of the 26th ACM SIGKDD
  International Conference on Knowledge Discovery \& Data Mining}, KDD '20,
  pages 1990--1998, New York, NY, USA. Association for Computing Machinery.

\bibitem{pattison2013conditional}
Pattison, P.~E., Robins, G.~L., Snijders, T. A.~B. {\&} Wang, P. (2013)
  Conditional estimation of exponential random graph models from snowball
  sampling designs. {\em Journal of Mathematical Psychology}, \textbf{57}(6),
  284--296.

\bibitem{pearson2006homophily}
Pearson, M., Steglich, C. {\&} Snijders, T. (2006)  Homophily and assimilation
  among sport-active adolescent substance users. {\em Connections},
  \textbf{27}(1), 47--63.

\bibitem{penrose2003random}
Penrose, M. (2003) {\em Random Geometric Graphs}.
Oxford University Press, Oxford, UK.

\bibitem{misspecification}
Ramsey, J.~B. (1969)  Tests for specification errors in classical linear
  least-squares analysis. {\em Journal of the Royal Statistical Society, Series
  B}, \textbf{31}(3), 350--371.

\bibitem{TrainRevelt97mixedlogit}
Revelt, D. {\&} Train, K. (1998)  {Mixed logit with repeated choices:
  {H}ouseholds' choices of appliance efficiency level}. {\em The Review of
  Economics and Statistics}, \textbf{80}(4), 647--657.

\bibitem{schmid2021generative}
Schmid, C.~S., Chen, T. H.~Y. {\&} Desmarais, B.~A. (2021)  Generative dynamics
  of Supreme Court citations: {A}nalysis with a new statistical network model.
  {\em Political Analysis}.
DOI: \url{10.1017/pan.2021.20}.

\bibitem{schmid2017exponential}
Schmid, C.~S. {\&} Desmarais, B.~A. (2017)  Exponential random graph models
  with big networks: {M}aximum pseudolikelihood estimation and the parametric
  bootstrap. {\em 2017 IEEE International Conference on Big Data (Big Data)},
  pages 116--121.

\bibitem{schweinberger2020exponential}
Schweinberger, M., Krivitsky, P.~N., Butts, C.~T. {\&} Stewart, J.~R. (2020)
  Exponential-family models of random graphs: inference in finite, super and
  infinite population scenarios. {\em Statistical Science}, \textbf{35}(4),
  627--662.

\bibitem{shalizi2013consistency}
Shalizi, C.~R. {\&} Rinaldo, A. (2013)  Consistency under sampling of
  exponential random graph models. {\em Annals of Statistics}, \textbf{41}(2),
  508--535.

\bibitem{siegel2009social}
Siegel, D.~A. (2009)  Social networks and collective action. {\em American
  Journal of Political Science}, \textbf{53}(1), 122--138.

\bibitem{Simon1955}
Simon, H.~A. (1955)  On a class of skew distribution functions. {\em Models of
  Man}, \textbf{42}, 145--164.

\bibitem{slyder2011citation}
Slyder, J.~B., Stein, B.~R., Sams, B.~S., Walker, D.~M., Jacob~Beale, B.,
  Feldhaus, J.~J. {\&} Copenheaver, C.~A. (2011)  Citation pattern and
  lifespan: {A} comparison of discipline, institution, and individual. {\em
  Scientometrics}, \textbf{89}(3), 955--966.

\bibitem{snijders2001statistical}
Snijders, T. A.~B. (2001)  The statistical evaluation of social network
  dynamics. {\em Sociological Methodology}, \textbf{31}(1), 361--395.

\bibitem{snijders2002markov}
Snijders, T. A.~B. (2002)  Markov chain {Monte Carlo} estimation of exponential
  random graph models. {\em Journal of Social Structure}, \textbf{3}, 2.

\bibitem{snijders2017stochastic}
Snijders, T. A.~B. (2017)  Stochastic actor-oriented models for network
  dynamics. {\em Annual Review of Statistics and Its Application},
  \textbf{4}(1), 343--363.

\bibitem{snijders2010introduction}
Snijders, T. A.~B., Van~de Bunt, G.~G. {\&} Steglich, C. E.~G. (2010)
  Introduction to stochastic actor-based models for network dynamics. {\em
  Social Networks}, \textbf{32}(1), 44--60.

\bibitem{stivala2020exponential}
Stivala, A., Robins, G. {\&} Lomi, A. (2020)  Exponential random graph model
  parameter estimation for very large directed networks. {\em PloS ONE},
  \textbf{15}(1), e0227804.

\bibitem{tomlinson2020learning}
Tomlinson, K. {\&} Benson, A.~R. (2021)  Learning Interpretable Feature Context
  Effects in Discrete Choice. In {\em Proceedings of the 27th ACM SIGKDD
  Conference on Knowledge Discovery \& Data Mining}, KDD '21, pages 1582--1592,
  New York, NY, USA. Association for Computing Machinery.

\bibitem{tomlinson2022graph}
Tomlinson, K. {\&} Benson, A.~R. (2022)  Graph-based methods for discrete
  choice. \textbf{arXiv:2205.11365}.

\bibitem{tomlinson2021choice}
Tomlinson, K., Ugander, J. {\&} Benson, A.~R. (2021)  Choice set confounding in
  discrete choice. In {\em Proceedings of the 27th ACM SIGKDD Conference on
  Knowledge Discovery \& Data Mining}, KDD '21, pages 1571--1581, New York, NY,
  USA. Association for Computing Machinery.

\bibitem{train1986qualitative}
Train, K. (1986) {\em Qualitative Choice Analysis: Theory, Econometrics, and an
  Application to Automobile Demand}.
MIT Press, Cambridge, MA, USA.

\bibitem{Train_Discrete_Book}
Train, K. (2009) {\em Discrete Choice Methods with Simulation}.
Cambridge University Press, Cambridge, UK.

\bibitem{valeeva2020duality}
Valeeva, D., Heemskerk, E.~M. {\&} Takes, F.~W. (2020)  The duality of firms
  and directors in board interlock networks: {A} relational event modeling
  approach. {\em Social Networks}, \textbf{62}, 68--79.

\bibitem{socialanalysisbook}
Wasserman, S. {\&} Faust, K. (1994) {\em Social Network Analysis: Methods and
  Applications}.
Cambridge University Press, Cambridge, UK.

\bibitem{wooldridge2016introductory}
Wooldridge, J.~M. (2016) {\em Introductory Econometrics: A Modern Approach}.
Cengage Learning, Boston, MA, USA, 6th edition.

\bibitem{Yamamoto2010AMR}
Yamamoto, T. (2014)  A multinomial response model for varying choice sets, with
  application to partially contested multiparty elections. Available at
  \url{http://web.mit.edu/teppei/www/research/dchoice.pdf}.

\bibitem{yeung2019statistical}
Yeung, F. (2019) {\em Statistical Revealed Preference Models for Bipartite
  Networks}.
PhD thesis, UCLA.
available at {\tt https://escholarship.org/uc/item/0fm6h8gm}.

\bibitem{Yule1925}
Yule, G.~U. (1925)  A mathematical theory of evolution, based on the
  conclusions of {Dr. J. C. Willis, F.R.S.}. {\em Philosophical Transactions of
  the Royal Society of London. Series B, Containing Papers of a Biological
  Character}, \textbf{213}, 21--87.

\end{thebibliography}



\newpage
\appendix

\section{Appendix: Additional Details of our Comparison Between the RC and MNL Models}\label{appendix1}


\subsection{Random Parameters}\label{sec:AppendixRandomParameters}

In Section \ref{subsec:Random-Choice-Parameters}, we compared the performance of the original MNL model and the RC model on a single network that we generated using using the method in Section \ref{subsec:SynEstimatiom}. The choice parameters are random for this network. We chose the network by constructing it uniformly at random from a set of 50 networks; we refer to the chosen network as the `sampled network'. In Section \ref{subsec:Random-Choice-Parameters}, we compared the MNL and RC models for this single network (rather than for the entire set) to simplify our exposition. We now provide summary statistics of our results from applying the MNL model to the set of 50 networks. The model specification for the MNL model is the same as in Sections \ref{subsec:SDatanadModel} and \ref{subsec:Random-Choice-Parameters}. We show our results in Table \ref{tab:AppendixRandomParameters}.

\begin{table}[H]
\caption{\textbf{{Summary statistics of the estimated values of the choice parameters in the MNL model for the set of 50 synthetic networks that we generate using the method in Section \ref{subsec:SynEstimatiom}.}}
The first column specifies the distribution parameters, the second column gives the mean estimated values of the
parameters, the third column gives the median estimated values of the
parameters, the fourth column gives the minimum estimated values of
the parameters, the fifth column gives the maximum estimated values
of the parameters, and the last column gives the standard errors
of the estimated values. 
}
\label{tab:AppendixRandomParameters}

\vspace{1cm}

\centering{}%
\begin{tabular}{p{1.5cm}  p{1.5cm}  p{1.5cm}  p{1.5cm}  p{1.5cm}  c}
\hline 
Parameter & Mean & Median & Minimum & Maximum & Standard Error\tabularnewline
\hline 
\hline 
$\beta_{1}$ & 1.845 & $1.844$ & $1.424$ & 2.310 & $0.227$\tabularnewline
\hline 
$\beta_{2}$ & 0.806 & $0.807$ & 0.573 & $0.966$ & $0.085$\tabularnewline
\hline 
\end{tabular}
\end{table}

The central values (both the mean and the median) of the parameter estimates across the set of 50 networks are similar in magnitude
to the estimated values of the sampled network, suggesting that the sampled network is representative of the set of networks. 
The small standard errors suggest that the results from Section \ref{subsec:Random-Choice-Parameters}
reflect most of the networks in the set.


\subsection{Deterministic Parameters}\label{AppendixDeterministicParameters}

In Section \ref{subsec:Deterministic-Choice-Parameters}, we showed results from applying the MNL and RC models on a single sampled network from the set of 50 networks that we generated using the original MNL model in Section \ref{subsec:SynMNLcomparison}. To simplify our exposition, we presented results for a single network that we chose uniformly at random from the set of networks; we refer to the chosen network as the `sampled network'.
We now provide summary statistics of the results of applying the MNL and RC models to the set of 50 networks that we generated using the MNL model.
The model specifications for the RC and MNL models are the same as in Section \ref{subsec:SDatanadModel}.
We show our results in Table \ref{tab:AppendixDeterministicParameters}.

\begin{table}[H]
\caption{\textbf{{Summary statistics of the estimated values of the choice parameters from the MNL and RC models for the set of 50 synthetic networks that we described in Section \ref{subsec:SDatanadModel}.}} The first column specifies the distribution parameters,
the second column gives the mean estimated values of the parameters,
the third column gives the median estimated values of the parameters,
the fourth column gives the minimum estimated values of the parameters,
the fifth column gives the maximum estimated values of the parameters,
and the last column gives the standard errors of the estimated
values. 
}
\label{tab:AppendixDeterministicParameters}

\vspace{1cm}

\centering{}%
\begin{tabular}{p{1.5cm}  p{1.5cm}  p{1.5cm}  p{1.5cm}  p{1.5cm}  c}
\hline \hline
Parameter& Mean & Median  & Minimum  & Maximum  & Standard Error\tabularnewline
\hline 
$\beta_{1}$ & $-1.010$ & $-1.014$ & $-1.324$ & $-0.754$ & $0.114$\tabularnewline
\hline 
$\beta_{2}$ & $3.035$ & $3.053$ & $2.716$ & $3.486$ & $0.142$\tabularnewline
\hline 
$m_{1}$ & $-1.020$ & $-1.020$ & $-1.341$ & $-0.773$ & $0.115$\tabularnewline
\hline 
$s_{1}$ & $0.053$ & $0.063$ & $-0.363$ & $0.495$ & $0.236$\tabularnewline
\hline 
$m_{2}$ & $1.122$ & $1.130$ & $1.000$ & $1.251$ & $0.049$\tabularnewline
\hline 
$s_{2}$ & $0.041$ & $0.028$ & $-0.101$ & $0.208$ & $0.079$\tabularnewline
\hline 
\end{tabular}
\end{table}

For all of the parameters, the central values (both the mean and the median) across the 50 networks in the set are similar in magnitude to the estimated values of the sampled
network, suggesting that the sampled network is {representative of the {set of networks}}. The small standard errors of $\beta_{1}$, $\beta_{2}$, $m_{1}$, and
$m_{2}$ support the results of Section \ref{subsec:Deterministic-Choice-Parameters} that the RC model gives accurate estimates of the parameters when we generate data using the RC model. However, the standard errors for $s_{1}$ and $s_{2}$ are relatively large. Given the large standard deviations, the strategy that we used in Section \ref{subsec:Deterministic-Choice-Parameters} suggests that the parameters are random. Specifically, for approximately $40$\% of the networks in our simulations, our approach suggests that the parameters are random.
Because the parameters are deterministic (by construction), this approach leads us to an incorrect conclusion about these networks.
We do not study this test in a formal statistical way in the present paper, and it is desirable to do so.


\end{document}